\documentclass[journal,11pt,draftclsnofoot,onecolumn,doublespace]{IEEEtran}

\usepackage{amsmath,amssymb,epsfig,psfrag,cite,subfigure}
\include{macros}
\usepackage{graphicx}
\usepackage{epstopdf}

\usepackage[detect-all,detect-weight=true]{siunitx}

\usepackage{nicefrac}

\usepackage{lipsum,multicol}

\usepackage{algorithm}
\usepackage{algpseudocode}

\usepackage{enumitem}
\usepackage{mwe}
\usepackage{epsfig,psfrag}
\usepackage{subfigure}
\usepackage{color}
\usepackage{url}
\usepackage{mathtools,xparse}

\usepackage{gensymb}

\usepackage{dsfont}

\usepackage[T1]{fontenc}
\usepackage[utf8]{inputenc}
\usepackage{authblk}

\allowdisplaybreaks

\DeclarePairedDelimiter{\abs}{\lvert}{\rvert}%
\DeclarePairedDelimiter{\absbig}{\Big\lvert}{\Big\rvert}%
\newcommand{\dd}{ \mathrm{d} }

\newcommand{\specificthanks}[1]{\@fnsymbol{#1}}

\newcommand{\rect}[2]{{ {\rm{rect}}_{#1}\left(#2\right) }}

\newcommand{\xconjt}{ x^{*}(t) }
\newcommand{\taumax}{ \tau_{\rm{max}} }

\newcommand{\ykn}{{ y_{k,n} }}
\newcommand{\wkn}{{ w_{k,n} }}
\newcommand{\nmax}{{ n_{\rm{max}} }}

\newcommand{\floor}[1]{\lfloor #1 \rfloor}
\newcommand{\Bmax}{ B_{\rm{s}} }

\newcommand{\tauhat}{{ \widehat{\tau} }}
\newcommand{\nuhat}{{ \widehat{\nu} }}

\newcommand{\Rd}{\Delta R}
\newcommand{\vd}{\Delta v}

\newcommand{\Badc}{ B_{\rm{adc}} }

\newcommand{\rmint}{ \rm{int} }

\newcommand{\gammaint}{{ \gamma_{\rm{int}} }}
\newcommand{\tauint}{{ \tau_{\rm{int}} }}
\newcommand{\rint}{{ R_{\rm{int}} }}

\newcommand{\xintkn}{{ x_{{\rm{int}},k,n} }}

\newcommand{\sintt}{ s_{\rmint} }

\newcommand{\xintt}{ x_{\rmint} }

\newcommand{\tildealpha}{{ \widetilde{\alpha} }}
\newcommand{\tildeT}{{ \widetilde{T} }}

\newcommand{\tildegamma}{{ \widetilde{\gamma} }}

\newcommand{\tildeb}{{ \widetilde{B} }}

\newcommand{\xmln}{ x_{m,\ell,n} }
\newcommand{\ymln}{ y_{m,\ell,n} }
\newcommand{\wmln}{ w_{m,\ell,n} }

\newcommand{\Tsym}{ T_{\rm{sym}} }
\newcommand{\Tcp}{ T_{\rm{cp}} }

\newcommand{\deltaf}{ \Delta f }

\newcommand{\Btot}{ B_{\rm{tot}} }
\newcommand{\Ttot}{ T_{\rm{tot}} }

\usepackage[normalem]{ulem}
\usepackage{hyperref}

\begin{document}

\title{Radar Interference Mitigation for\\ Automated Driving}
\author{
{Canan Aydogdu},
{Gisela K. Carvajal},
{Olof Eriksson},
{Hans Hellsten},
{Hans Herbertsson},
{Musa Furkan Keskin},
{Emil Nilsson},
{Mats Rydström},
{Karl Vanäs},
{Henk Wymeersch}
\thanks{Canan Aydogdu, Musa Furkan Keskin, and Henk Wymeersch are with the Department of Electrical Engineering, Chalmers University of Technology, 41296 Gothenburg, Sweden, Email: canan.aydogdu@chalmers.se, Gisela K. Carvajal is with QAMCOM Research, 41285 Gothenburg, Sweden. 
Karl Vanäs is with Volvo Car Corporation, 40531, Gothenburg, Sweden.  Hans Herbertsson, Olof Eriksson and Mats Rydström are with Veoneer, 44737, Vårgårda, Sweden. Hans Hellsten and Emil Nilsson are with Halmstad University, Sweden. Hans Hellsten is also with SAAB. This research was supported by Vinnova grant 2018-01929. 
}}


\maketitle

\begin{abstract}
Autonomous driving relies on a variety of sensors, especially on radars, which have unique robustness under heavy rain/fog/snow and poor light conditions. 
With the rapid increase of the amount of radars used on modern vehicles, where most radars operate in the same frequency band, the risk of radar interference becomes a compelling issue. 
This article analyses automotive radar interference and proposes several new approaches, which combine industrial and academic expertise, toward the path of interference-free autonomous driving.
\end{abstract}

\markboth{IEEE Signal Processing Magazine, Special Issue on Autonomous Driving, \today}%
{Shell \MakeLowercase{\textit{et al.}}: Bare Demo of IEEEtran.cls for IEEE Journals}


\section*{Introduction and Motivation}
Radar is becoming the standard equipment in all modern cars, supporting, e.g., cruise control and collision avoidance 
in most weather conditions whilst providing high-resolution detections on the order of centimeters in the millimeter-wave (mmWave) band. The next generation of Advanced Driver Assistance (ADAS) and Autonomous Drive (AD) vehicles will have a multitude of radars covering multiple safety and comfort applications like crash-avoidance, self-parking, in-cabin monitoring, cooperative driving, collective situational awareness and more. 

\begin{figure}
	\centering
	
	\includegraphics[width=0.7\columnwidth]{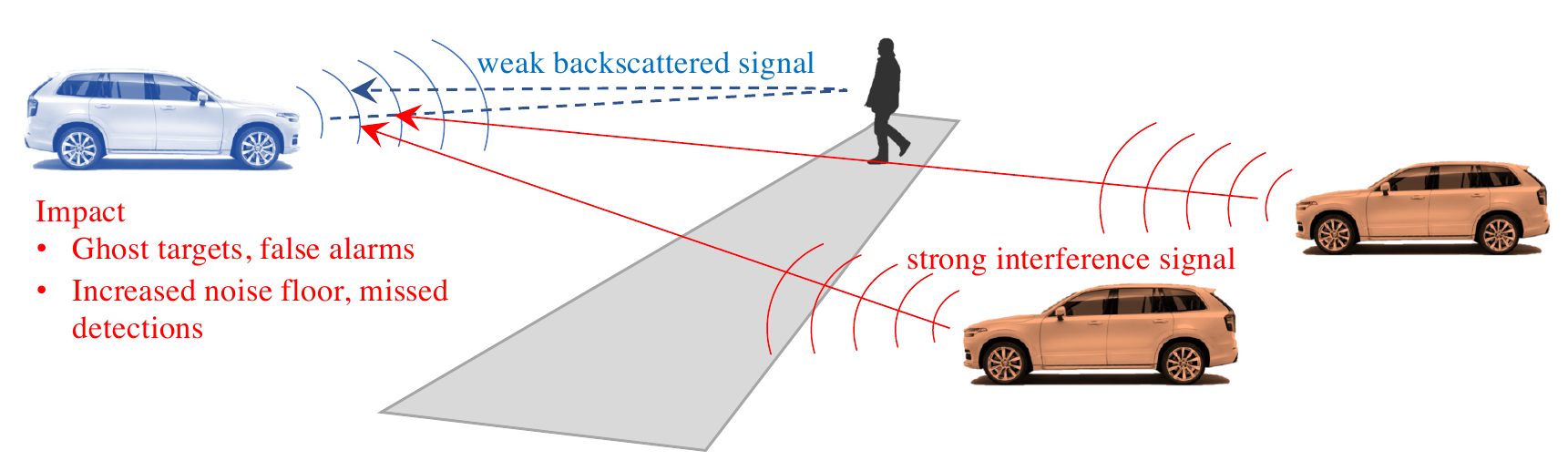}
	\caption{Interference is generally much stronger than the desired radar signal, due to the one-way propagation. Interference increases with more interfering radars and leads to false alarms and missed detections. }
	\label{fig_interferenceImpact}
	\vspace{-0.5cm}
\end{figure}

Since automotive radar transmissions are uncoordinated, there is a non-negligible probability of interference among vehicles, as shown in Fig.~\ref{fig_interferenceImpact}. While current automotive radars are already impacted by interference to some extent, it is today unlikely to get issues noticeable to the customer as the state-of-the-art automotive radars are continuously updated and improved on multiple system levels. However, the mutual interference problem is expected to become more challenging, unless properly handled, as more vehicles are equipped with a larger number of radars providing $360 \degree$ situational awareness at various distances to enable more advanced future ADAS and AD functionalities. This is evidenced by multiple international studies, such as in the EU MOSARIM project \cite{MosarimFinalReport} and the more recent IMIKO RADAR project.  
All major players in the automotive sensor market, like Volvo and Veoneer, are involved in activities studying the next generation of “interference-free radars”. This includes, for example, enhancing models to see the impact of a larger density of radars, simulating new interference scenarios, investigating different medium access control (MAC) models and methods to coordinate radar transceivers, both decentralized and centralized. At this point, 
the automotive industry is ready to consider novel designs and approaches, which might impact standardization bodies before new frequency spectrum is made available in the higher RF bands.


Signal processing can provide ways to reduce or mitigate interference, both at the raw signal level as well as at the post-detection/target tracking level. The particular properties and requirements of automotive radar impose significant challenges in terms of signal processing. This includes combination of radar and communication waveforms, which brings up further possibilities regarding ultra-reliable low-latency communications (URLLC) in vehicular ad-hoc networks (VANETs). Hence, it is timely to review what has been done, 
what are reasonable approaches, and what the future holds. 

The focus of this article is on frequency modulated continuous wave (FMCW)  radar, since it is the most common and robust
automotive radar. 
We provide an analysis of the impact of interference in FMCW
both quantitatively and qualitatively, in terms of their probability, severity, and effects. Then, we cover different ways to 
mitigate interference, ranging from changing FMCW parameters, to new signal structures, and explicit coordination between vehicles. We also study new techniques that are potentially more robust towards interference, including stepped-frequency orthogonal frequency division multiplexing (OFDM).  
Finally, we describe what we believe will be the long-term evolution of automotive radar and its relation to mobile communication. 

\section*{Automotive Radar}
\subsection*{History and Future of Automotive Radar}

Radar has been used in automotive applications for a long time. Already in 1949, unfortunate car drivers were issued speeding tickets based on speed measurements obtained from the radar speed gun, recently invented by John L. Barker Sr.~\cite{patent_radargun}.  However, on-board automotive radar was not made commercially available until 1999, when it was introduced for collision warning and automatic cruise control (ACC). See \cite{meinel_dickmann} for an early history of automotive radar with some entertaining vintage photography. 
 Over the years, there has been a strong push to increase the integration level of millimeter wave electronics used for automotive radar and industrial radar sensors. The early discrete hardware designs have been replaced by a few chips in III-V-materials, and now CMOS single chip solutions are available. With CMOS technology comes the ability to fully integrate analog and digital electronics, making very advanced protocols and detection schemes possible at low cost and low power. 
Consequently, radar is becoming more and more common for supporting various automotive applications. ADAS systems based on radar are today standard equipment in most new vehicles. Vehicles capable of some level of AD are also foreseen to rely, at least to some degree, on radar systems for monitoring vehicle surroundings. The number of radar transceivers operating throughout the traffic environment is foreseen to increase rapidly over the coming years. 
As the number of radar transceivers in the traffic infrastructure increases, radar interference is also expected to increase. Today most radar transmissions are uncoordinated, meaning that there is no a priori agreement on who is allowed to transmit and when.  A number of recent studies have indicated the interference situations which are likely to arise as the automotive radar transceiver market penetration increases \cite{RadarCongestionStudy,MosarimFinalReport}. 
FMCW waveforms can, up to a point, relatively easily be repaired in case it is intermittently corrupted by interference \cite{patent_repair_method}, which is why they are still operational. 

Future radar systems are expected to occupy frequency bands higher and higher up in frequency. Transceivers operating around 77 GHz are available today, and transceivers operating at carrier frequencies beyond 100 GHz are expected
. Frequencies as high as 300 GHz and beyond are being considered for some applications, such as synthetic aperture radar mapping. Operation at such high frequencies bring the obvious benefits of improved miniaturization, but also presents challenges in terms of hardware complexity and signal attenuation. Moreover, interference-free operation will require  radar transmission standardization. A standardized transmission scheduling system resembling today's cellular communication system would present a solution to the interference problem, but it is not without challenges, both technical and political. 

\subsection*{Basics of FMCW Radar}


In a general FMCW radar a frequency sweep, a chirp, is generated by a voltage-controlled oscillator (VCO) controlled by a digital synthesizer. The generated chirp signal is split and sent into two different signal paths: one path is directed to the transmitter antenna (TX), while the other path is directed to the mixer correlator. Before the chirp is sent out on the TX antenna it passes a power amplifier (PA) boosting the transmitted energy. The transmit waveform of an FMCW radar with $K$ consecutive linear frequency modulated (LFM) chirps (or sweeps) can be expressed as \cite{fmcw_92,patole2017automotive}
\begin{equation} \label{eq_base_transmit}
s(t) =   \sum_{k=0}^{K-1} x(t - kT) 
\end{equation}
where the individual chirps are given by
\begin{equation}\label{eq_base2}
x(t) =  e^{j \varphi(t)} \rect{T}{t}, ~ \varphi(t) = 2\pi (f_c t + 0.5 \alpha t^2)~.
\end{equation}
Here, $\alpha = {B}/{T}$ is the chirp slope, $B$ denotes the sweep bandwidth, $T$ represents the chirp duration, $f_c$ is the carrier frequency, and $\rect{T}{t} $ is square pulse of duration $T$ with amplitude $1$. 
The received reflected signal from a target is very weak due to the two‑way free space propagation path loss and losses incurred during reflection, and thus needs to be amplified with a low noise amplifier (LNA) to maintain an acceptable signal-to-noise-ratio (SNR). The amplified signal from the target reflection is correlated with the transmitter signal in the mixer correlator, also called dechirping. The low pass filter at the output of the correlating mixer offers some interference rejection. Round-trip delay and Doppler shift caused by the relative velocity of the target shifts the frequency of the received signal compared to the transmitted signal. As a result, the mixer creates a beat signal that will pass through a low-pass filter and be digitized, yielding delay and Doppler estimates after matched filtering. In modern automotive radars, it is also possible to estimate azimuth and elevation of targets using multiple antennas. 



\begin{figure}
	\centering
	\includegraphics[width=0.6\columnwidth]{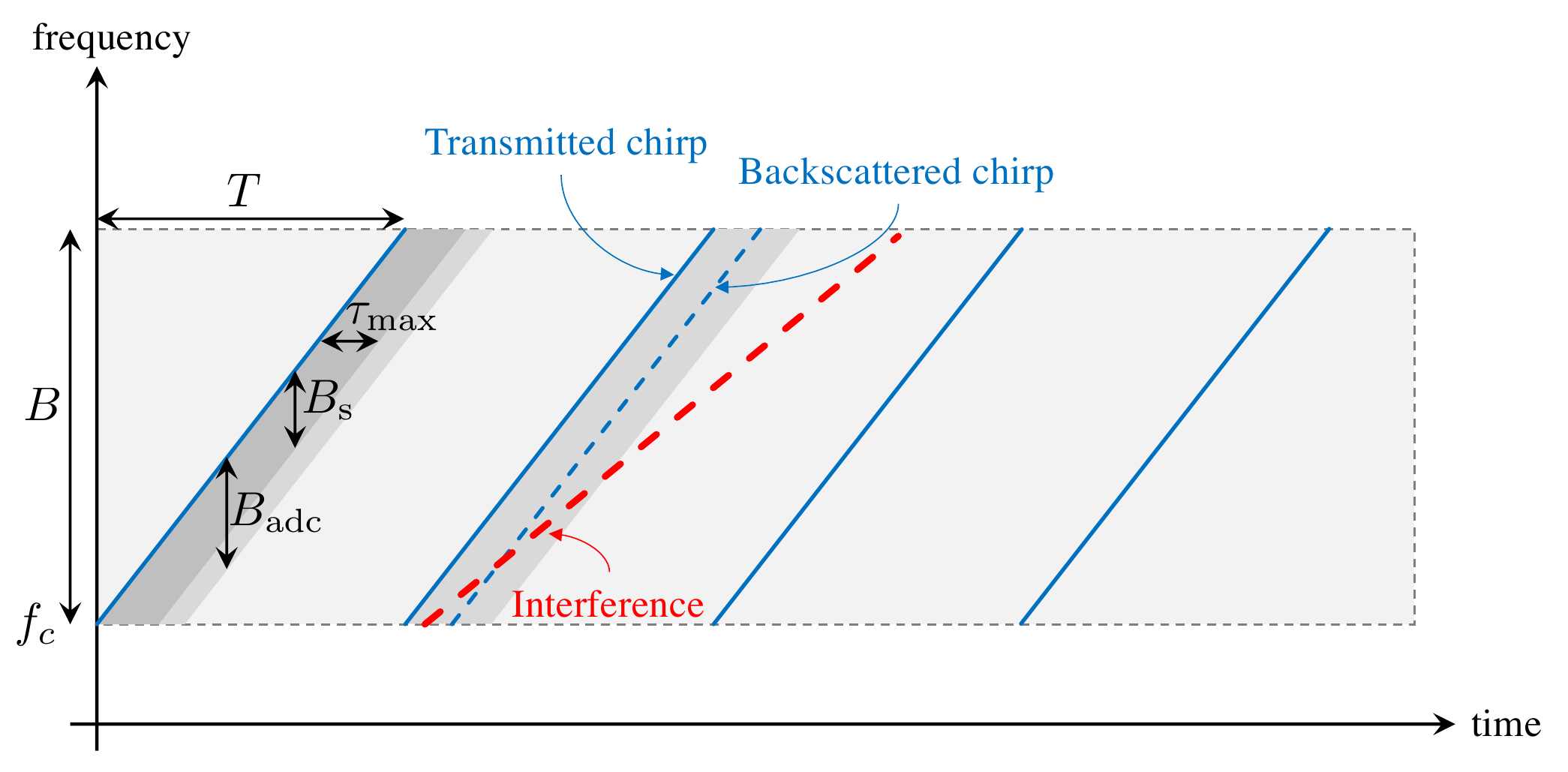}
	\vspace{-0.3cm}
	\caption{Four consecutive chirps in time-frequency representation. Several key notations are included. A backscattered signal and  an interfering signal is shown at the second chirp.}
	\label{fig_chirpDef}
	\vspace{-0.5cm}
\end{figure}

\subsection*{FMCW Radar Signal Processing Chain}


Suppose there exists a single target of interest acting as a point scatterer, characterized by a complex channel gain $\gamma$ (including the effects of path loss, antenna gain and radar cross section), an (initial) round-trip propagation delay $\tau = 2 R /c$ and a normalized relative Doppler shift $\nu = 2 v/c$, where $R$ and $v$ denote, respectively, the distance and relative radial velocity between the radar and target, and $c$ is the speed of wave propagation.
The received backscattered signal is now processed in three stages. 
\subsubsection*{Stage 1: Dechirping}
Under the stop-and-hop assumption \cite[Ch.~2.6.2]{richards2005fundamentals}, 
the $k$th chirp of the received signal 
is given by
\begin{equation} \label{eq_rt_fmcw2}
r_k(t) = \gamma \, x( t + (t + kT)\nu - \tau ) + w_k(t) 
\end{equation}
where $0 \leq t \leq T$ denotes the time relative to the beginning of the $k$th chirp, where $w_k(t)$ is measurement noise. 
To obtain the \textit{beat signal} at the intermediate frequency (IF), the received signal $r_k(t)$ in \eqref{eq_rt_fmcw2} is \textit{dechirped} through conjugate mixing with the transmitted signal $x(t)$ in \eqref{eq_base2}\footnote{Here, we ignore the terms whose total phase progression over a coherent processing interval (CPI) of $K$ chirps is smaller than $\pi/4$ for typical automotive FMCW settings \cite[Ch.~2.6.3]{richards2005fundamentals}.}:
\begin{align} \label{eq_rt_fmcw3}
& y_k(t) = r_k(t) \xconjt = \gamma \, e^{j 2\pi f_c \nu (t+kT)} e^{-j 2\pi \alpha \tau t} \rect{T}{{t - \tau}}  + w_k(t) \xconjt.
\end{align}
Let $\taumax$ denote the round-trip delay (see Fig.~\ref{fig_chirpDef}) corresponding to a maximum target range of interest (i.e., $\taumax \geq \tau$), which is related to the \textit{radar bandwidth of interest} $\Bmax$ as\footnote{The ADC bandwidth $\Badc \geq \Bmax$ imposes a limit on $\Bmax$, and, thus, the maximum detectable range $\taumax$.} $\taumax = \Bmax/\alpha$. After low-pass filtering the beat signal in \eqref{eq_rt_fmcw3} with bandwidth $\Bmax$, sampling with a period of $T_s$ for $\taumax \leq t \leq T$, we rearrange into a slow-time-fast-time data matrix, where the $k$th row contains the samples of the $k$th chirp (fast time), while the $n$th column contains the $n$th sample of each chirp (slow time). In other words, 
we have
\begin{align}\label{eq_ykn}
\ykn &= \gamma \, e^{j 2\pi (- \alpha \tau + f_c \nu) n T_s } e^{j 2\pi f_c \nu k T} + \wkn
\end{align} 
for $k = 0, \ldots, K-1$ and $n = \nmax, \ldots, N-1 $, $\nmax = \floor{\taumax/T_s}$, $N = \floor{T/T_s} + 1$, and $\wkn$ are independent and identically distributed (i.i.d.) complex Gaussian noise samples with variance $\sigma^2$.

\subsubsection*{Stage 2: Target Range-Velocity Estimation}
To provide an estimate of target range and velocity, two-dimensional (2-D) discrete Fourier transform (DFT) can be applied across slow-time and fast-time dimensions of the beat signal in \eqref{eq_ykn}, which yields the FMCW delay-Doppler spectrum evaluated at a given delay-Doppler pair $(\tauhat, \nuhat)$:
\begin{align}
z(\tauhat, \nuhat) &= \sum_{k=0}^{K-1} \sum_{n=\nmax}^{N-1} \ykn e^{j 2 \pi \alpha \tauhat n T_s} e^{-j 2\pi f_c \nuhat k T} ~. \label{eq_range_doppler}
\end{align}
The periodogram $\abs{z(\tauhat, \nuhat)}^2$ corresponding to \eqref{eq_range_doppler} yields a dominant target peak at $(\tauhat,\nuhat) = (\tau - f_c \nu/\alpha, \nu)$, which can be recovered using, for example, constant false alarm rate (CFAR) detectors \cite[Ch.~6]{richards2005fundamentals}. The peak value of $\abs{z(\tauhat, \nuhat)}$ is proportional to the processing gain $G_{\text{p}}=K(N-\nmax)$. 
This frequency identification method is referred to as the \textit{periodogram spectral estimator} \cite[Ch.~2.2.1]{stoica2005spectral}. Here, the shift $f_c \nu/\alpha$ in delay stems from \textit{range-Doppler coupling} inherent in the FMCW waveform \cite[Ch.~4.6.4]{richards2005fundamentals}. To compensate for the coupling effect, the Doppler-dependent term $f_c \nu/\alpha$ can be added back to the delay estimate after delay-Doppler retrieval.


When there are multiple objects in the radar field of view,  \eqref{eq_range_doppler} will have multiple peaks. 
In order to distinguish the different objects
, each object pair must be separated by a certain minimum gap in delay and Doppler domains, which is determined by the radar \textit{resolution}: the range and velocity resolution of an FMCW radar can be derived from \eqref{eq_range_doppler} as (assuming $\nmax \ll N$)
$\Rd = {c}/{(2 B)}$ and  $\vd = {\lambda_c}/{(2 KT)}$, where $\lambda_c$ is the carrier wavelength \cite[Ch.~2.4]{stoica2005spectral}.
Therefore, higher sweep bandwidth leads to better range resolution, while longer CPI duration means improved velocity resolution.

\subsubsection*{Stage 3: Tracking Filter}
At the final stage of the signal processing chain, the delay-Doppler detections $\{(\tau_p,\nu_p)\}_{p=0}^{P-1}$  (along with the corresponding azimuth-elevation pairs in the case of multiple antennas \cite{patole2017automotive}) are fed to a data association and tracking filter to provide filtered three-dimensional (3-D) positions and velocities of surrounding objects. Here, $P$ denotes the number of targets seen during one scan.

\section*{Is Interference Really a Problem?}
In this section, we provide a theoretical analysis for the impact of interference on the radar signal processing. We start with studying a single link and then extend to a network of vehicles on a multi-lane highway, in order to assess the impact of interference as a function of the vehicle and radar density, as well as the deployment  scenario. Our focus will be on direct interference from one radar to another. Indirect interference (i.e., scattered on objects) will be weaker and is ignored for conceptual simplicity. 

\subsection*{Single Link Interference}
 
The interfering radar employs the FMCW waveform $\sintt(t) = \sum_{k=0}^{K-1} \xintt(t - k\tildeT)$ where
$\xintt(t) =  e^{j 2\pi (f_c t + 0.5 \tildealpha t^2)} \rect{\tildeT}{t} $, 
while the victim radar utilizes the same waveform as specified in \eqref{eq_base_transmit} and \eqref{eq_base2}. Here,  $\tildealpha = {\tildeb}/{\tildeT}$, $\tildeb$ and $\tildeT$ denote, respectively, the chirp slope, sweep bandwidth and chirp duration of the interfering radar. The samples 
\eqref{eq_ykn} then become 
\begin{align}\nonumber
\ykn &= \gamma \, e^{j 2\pi (- \alpha \tau + f_c \nu) n T_s } e^{j 2\pi f_c \nu k T} + \gammaint \xintkn+ \wkn ~.
\end{align} 
Interference is generally much stronger than the desired back-scattered signal, as they are governed by the Friis free space propagation equation and the radar equation respectively \cite{stochastic_interference_its_2017}:
\begin{align}
 |\gamma_{\rm{int}}|^2 &=P\frac{G_{\text{trx}}\lambda^{2}}{(4\pi)^{2}r^{2}}~, \label{eq:interferencePower} \\
 |\gamma|^2& =P\frac{G_{\text{trx}}\sigma\lambda^{2}}{(4\pi)^{3}d^{4}}~, \label{eq:usefulPower}
\end{align}
where $r$ is the distance between the interferer and the victim radar, $d$ is the distance between the radar and the target, $P$ is the transmit power, $G_{\text{trx}}$ is the combined transmit and receive antenna gain, $\sigma$ is the radar cross section of the target. Hence, for similar $d$ and $r$ and typical values of $\sigma$, $|\gamma_{\rm{int}}|^2 \gg |\gamma|^2$. 
The nature of the interference depends on the \emph{total interference power} (i.e., the aggregate power of the interference samples), and the \emph{level of coherence} between victim and interfering radar \cite{goppelt2010automotive}. 

The total interference power depends on the statistics of the samples $\xintkn$, which is a function of the radar waveform parameters and signal delays. The samples satisfy $|\xintkn|^2\in \{0,1\}$, depending on whether or not the interference signal at time $(k,n)$ is in the bandwidth of interest of the victim radar. 
Hence, the overall power of the interference is  $\mathbb{E}\{\sum_{k,n}|\gamma_{\rm{int}}|^2|\xintkn|^2 \} = f |\gamma_{\rm{int}}|^2 G_{\text{p}}$, where 
$G_{\text{p}}$ is the radar processing gain and $f$ is the average interference probability. 

How this total interference power manifests itself depends on the radar parameters, and interference can be classified as coherent, incoherent, or partially coherent \cite{p2p_int_fmcw_2018}.
\textit{Coherent interference} occurs when the interferer uses the same parameters ($\alpha,T,B$) as the victim radar. In that case, the interfering radar signal leaking into the bandwidth of interest $\Bmax$ of the victim radar (i.e., $f=1$) leads to a \textit{ghost target}, a peak in the delay-Doppler spectrum with very high power \cite{FMCW_int_2011}. Ghost targets lead to false detections, which in turn may cause incorrect behavior of safety systems.  
\textit{Incoherent interference} occurs when the samples $\xintkn$ are independent random variables, due to  the interferer using very different waveform parameters (e.g., different chirp pattern) so that the total interference power $f |\gamma_{\rm{int}}|^2 G_{\text{p}}$
 ends up as an increased noise floor. 
Noise floor increase resulting from interference can lead to more severe degradation in detection performance than an equivalent increase in thermal noise floor due to the side-lobes of the 
interference spectrum \cite{p2p_int_fmcw_2018}.
In between these two extreme cases, a slight mismatch in chirp slope or chirp duration or in the presence of phase noise (\textit{partially coherent interference}) causes the energy of the ghost target peak (which occurs due to coherent interference) to spread over the delay-Doppler domain. 
%

\begin{figure}
	\centering
	\vspace{-0.5cm}
	\includegraphics[width=0.5\linewidth]{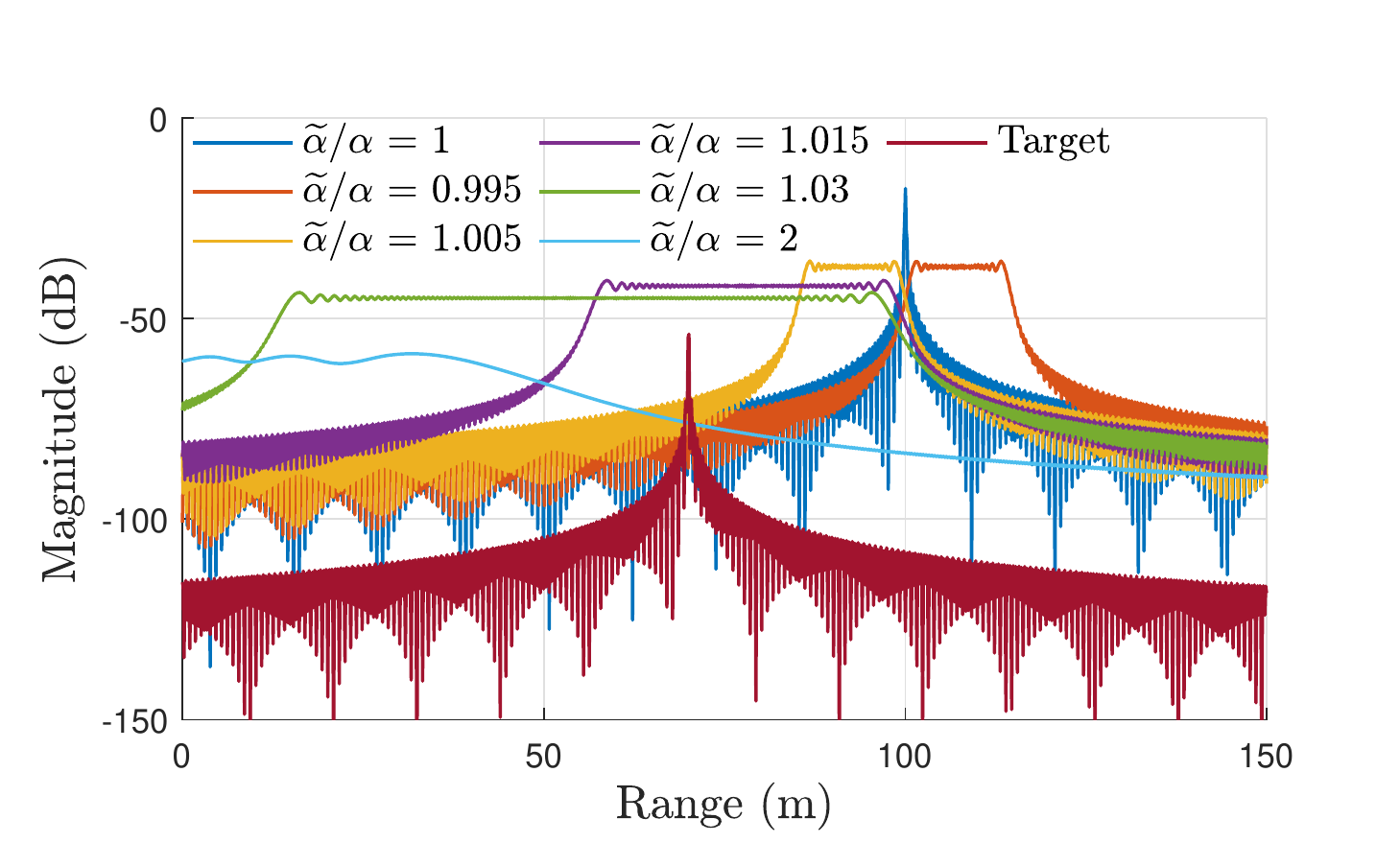}
	\vspace{-1cm}
	\caption{FMCW range profiles in the presence of an interfering radar for various values of chirp slope ratios, where $\tildealpha$ and $\alpha$ represent the chirp slopes of the interfering radar and the victim radar, respectively. The victim radar waveform parameters are $f_c = 77 \, \rm{GHz}$, $B = 1 \, \rm{GHz}$, $T = 20 \, \mu\rm{s}$. The interfering radar has identical chirp duration $\tildeT = 20 \, \mu\rm{s}$, but has a varying sweep bandwidth $\tildeb$ (thus, $\tildealpha$). The interference signal has a one-way propagation delay $\tauint$ corresponding to a range of $\rint = 100 \, \rm{m}$, while the desired target is located at $R = 70 \, \rm{m}$. Due to increased noise floor, the target may not be detected depending on its range and the chirp slope mismatch between the victim and interfering radars.}
	\label{fig_multiple_bws}
	\vspace{-0.5cm}
\end{figure}

To illustrate how a ghost target is spread out depending on the relative waveform parameters, Fig.~\ref{fig_multiple_bws} shows the fast time FFT output, i.e., the range FFT, corresponding to an interfering radar signal as a function of distance. The larger the difference in the chirp slope, the more the interference is spread out, due to the decrease in coherence. This affects the detection of targets in various ways. Incoherent interference may hinder the detection of low RCS targets (pedestrians, cyclists) over a large fraction of the delay-Doppler domain, whereas a (partially) coherent interference can mask even high RCS targets (vehicles) but in a smaller fraction of the delay-Doppler domain.

\begin{figure}
 	\centering
 	\vspace{-0.6cm}
 	\includegraphics[width=0.6\linewidth]{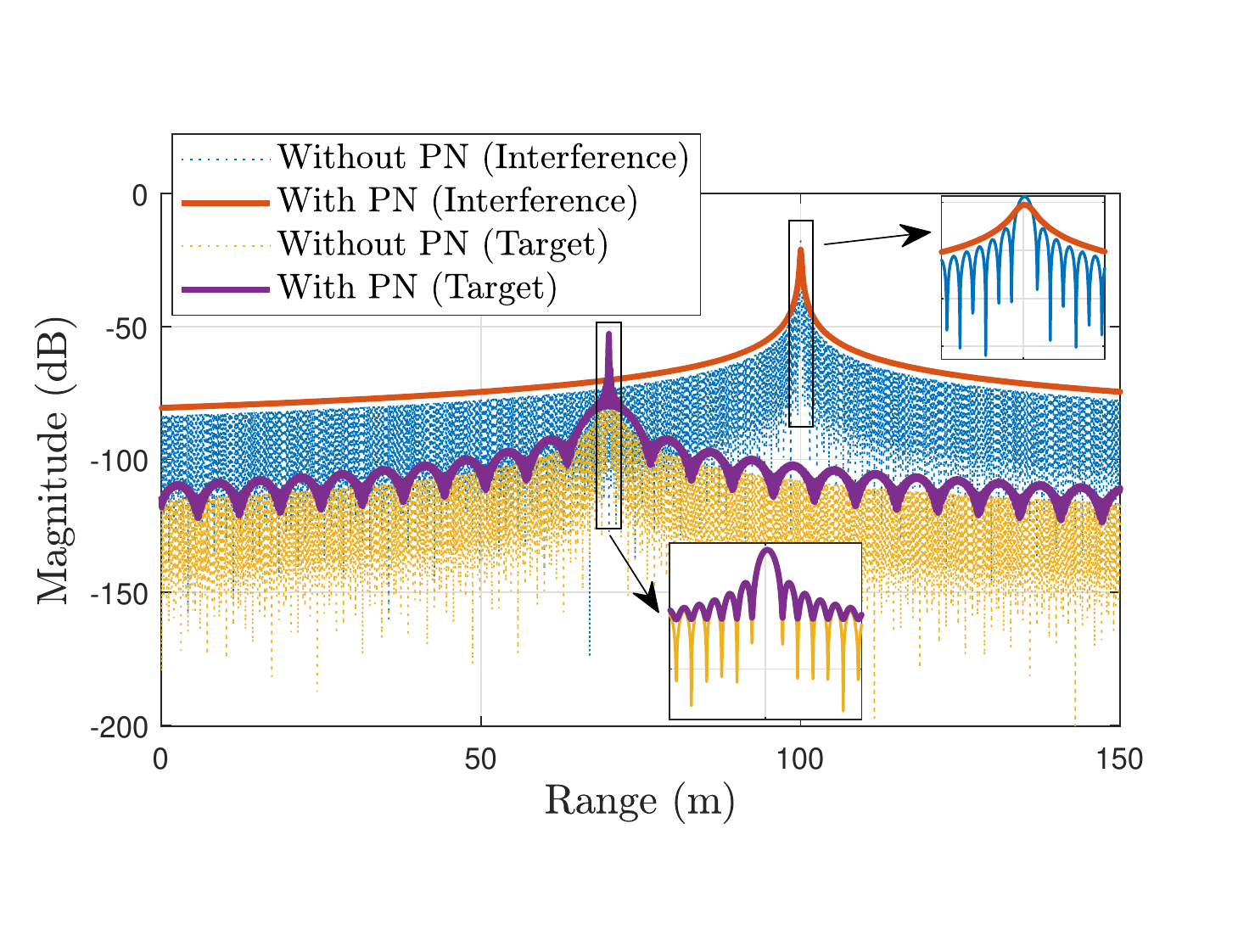}
 	\vspace{-1.5cm}
 	\caption{Range profiles of a victim FMCW radar in the presence of an interfering radar with identical chirp parameters (coherent interference), where both radars' oscillators have phase noise (PN) processes (the profiles without PN are also shown for comparison). The same parameters as in Fig.~\ref{fig_multiple_bws} are used with $\tildealpha/\alpha = 1$. Perfect range decorrelation of the interfering signal with the victim radar signal (due to independent phase noise processes at the victim and interfering radars) makes the spectral smearing effect more pronounced in the interference profile than in the target profile \cite{Range_Correlation_93}.}
 	\label{fig_phase_noise}
 	\vspace{-0.2in}
 \end{figure}

 In practice, oscillators in FMCW radars do not have an ideal, impulse-like radio-frequency (RF) spectrum due to phase and frequency instabilities \cite{PN_FMCW_2019_TAES}. In Fig.~\ref{fig_phase_noise}, we demonstrate the effect of oscillator \textit{phase noise} on the averaged range response\footnote{Range spectra are derived by computing the range FFTs of signal and interference powers averaged over the randomness of phase noise, which is modeled as a zero-mean wide sense stationary (WSS) random process under the assumption of white frequency modulated (FM) phase noise in the oscillator \cite[Sec.~V]{PN_2006}.} of a victim FMCW radar when the oscillators of both the victim and interfering radars are subject to phase noise processes with parameters $L_p = -70 \, \rm{dBc/Hz}$ (pedestal height) at $W_p = 200 \, \rm{kHz}$ (pedestal width) \cite{Demir_PN_2006,PN_FMCW_2019_TAES}. As observed from the figure, the oscillator phase noise induces spectral smearing of target and interference profiles, thereby causing loss of details in the spectrum, which deteriorates detection performance and leads to masking of weak targets. 

\subsection*{Network Interference}

\begin{figure}[h]
	\centering
	\includegraphics[width=0.6\linewidth]{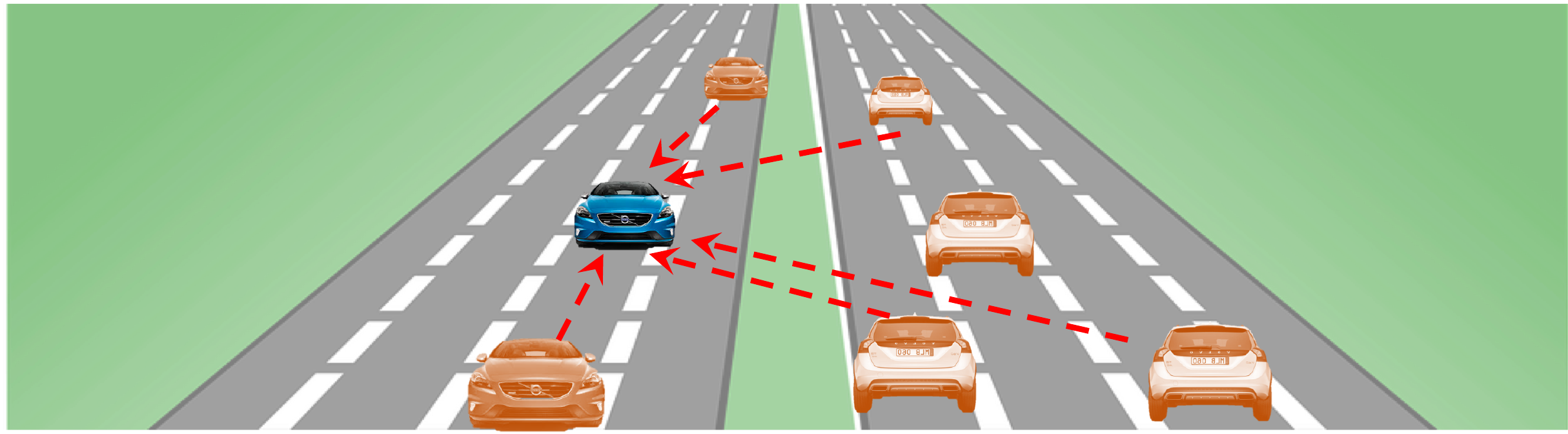}

	\caption{Example of network interference in a six-lane highway. Interference from interfering radars is aggregated and depends on the properties of the individual radars as well as the placement of vehicles on the road. 
	}
	\label{fig_HighwayScenario}
\end{figure}

\begin{figure}
	\centering
	\includegraphics[width=0.5\linewidth]{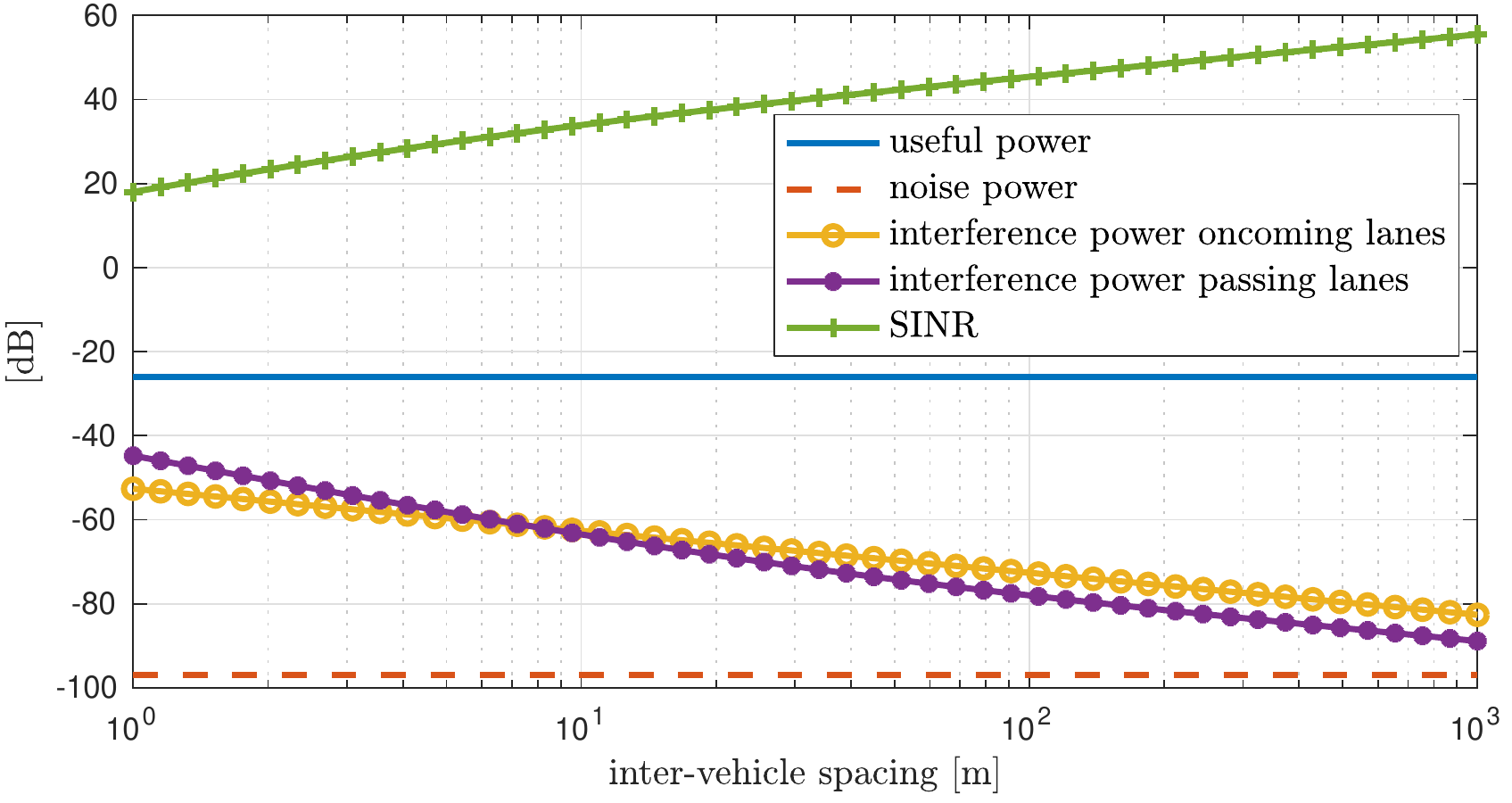}
    \vspace{-0.5cm}
	\caption{Incoherent network interference in a 6-lane highway as a function of the average vehicle spacing. The distance to the target is 150 m, 10 us chirp time, $\sigma = 10~\text{m}^2$, 10 dBm transmit power. 1 GHz bandwidth, $T= 30$ us, 150 m maximum range, 50 MHz ADC bandwidth, 10 dBm transmit power (same for front-end and back-end),  30 degree FOV in forward direction, 90 degree in backward direction, 20 \% duty cycle, 100 $\mu$s  frame duration. 
	}
	\label{fig_networkInterference}
	\vspace{-0.5cm}
\end{figure}

The above interference analysis can be extended to a complete network, for instance on a multi-lane road, by employing a stochastic geometry approach \cite{stochastic_interference_its_2017}. 
As shown in Fig.~\ref{fig_HighwayScenario}, consider a victim radar surrounded by $L$ lanes of traffic, with lane separation $R$, each modeled as a one-dimensional Poisson point process  (PPP) $\Phi(\mathbf{x})$ with intensity $1/\Delta$ (so $\Delta$ is the expected distance between vehicles and $\mathbf{x}$ is a vehicle location along a road) \cite{RadarCongestionStudy}. 
Radars (here one per side of the vehicle) are incoherent and can have different chirp durations, but otherwise share the same bandwidth $B$, duty cycle $u\in[0,1]$ (i.e., the fraction of time the radar is transmitting), and field of view.\footnote{We recall that for antennas with narrow field of view (FOV), the antenna gain is $G\approx 4\pi/(\phi\theta))$,
where $\phi$ is the beam-width in the elevation domain and $\theta$ the beam-width in the azimuth domain. This means that a radar with 1 degree elevation beam-width and 30 degrees azimuth beam-width will have a gain of approximately 22 dBi.} The expected value of the interference probability $f$ is easily found to be\footnote{Here we made use of the following asymptotic results. When 
$\tildealpha=\alpha$ then the probability of interference is $u\Bmax/B$ and the interference lasts an entire chirp duration. When $\tildealpha \ll \alpha$, then the probability of interference is
$u$ and the duration is $\tildealpha\taumax/(\alpha -\tildealpha)$. When $\tildealpha \gg \alpha$, then there are $\tildealpha/\alpha$ simultaneous interferers, each lasting for a duration $\alpha \taumax/(\tildealpha-\alpha)$. } $f= u \alpha \taumax/B$.
Hence, the aggregate interference seen by the victim radar due to interference from lane $\ell \in \mathbb{Z}$ (indexed with reference to the victim radar) is 
\begin{align}
I_{\text{p}}(\ell)=\sum_{\mathbf{x}\in\Phi(\mathbf{x}) \cap \mathbf{x}\in \text{FOV}} {P}f\frac{G_{\text{trx}}\lambda^{2}}{(4\pi)^{2}r^{2}(\mathbf{x})},    
\end{align}
where $r(\mathbf{x})=\sqrt{\ell^{2}R^{2}+x^{2}}$, where $x$ is the
1-dimensional position along the road, ranging from $\frac{\ell R}{\tan\theta_{\text{p}}/2}$
to $+\infty$. Here, $\theta_{\text{p}}$ is the minimum of the forward and backward field of view. Hence (with slight abuse of notation), the interference averaged over the locations of the interferers is
\begin{align}
\mathbb{E}\{ I_{\text{p}}(\ell)\}
  ={P}\frac{G_{\text{trx}}\lambda^{2}}{(4\pi)^{2}}\frac{f}{\Delta}\int_{\frac{\ell R}{\tan\theta_{\text{p}}/2}}^{+\infty}\frac{1}{\ell^{2}R^{2}+x^{2}}\text{d}x
  ={P}\frac{G_{\text{trx}}\lambda^{2}}{(4\pi)^{2}}\frac{f}{\Delta}\frac{1}{\ell R}\frac{\theta_{\text{p}}}{2}~,
\end{align}
while for $\ell=0$, $r(\mathbf{x})=x$, where we need a certain safety
margin to avoid singularities, we set $x$ from $\Delta$ to
$+\infty$, leading to $I_{\text{p}}(\ell=0)={P}\frac{G_{\text{trx}}\lambda^{2}}{(4\pi)^{2}}f\frac{1}{\Delta^{2}}$.  
An example of network interference of a six-lane highway is shown in Fig.~\ref{fig_networkInterference}, as a function of the average inter-vehicle spacing $\Delta$ for a vehicle target 150 meters away with RCS of $10~\text{m}^2$. The analytical result shows the impact of interference of nearby vehicles, leading to orders of magnitude reduction of the SINR. For small $\Delta$, interference is larger from passing lanes, while for large $\Delta$, oncoming traffic dominates. We also observe that even though interference power can be large, the factor $f$ reduces its impact significantly. In the example $f\approx 0.01$, leading to 20 dB reduction in interference. In this example, the target can be detected in spite of the incoherent network interference, while  a pedestrian target with smaller RCS (such as $0.1-1~\text{m}^2$) further away than $50~\text{m}$ would be hard to detect.

\subsection*{Intermediate Conclusion}
From the above analyses, we found that interference can manifest itself in different ways, and can increase both missed detections and false alarms. Due to the nature of the FMCW signals, there is a natural robustness to interference. Both the total received interference power and the mutual coherence between victim and interfering radar play an important role. 
As a rule of thumb, the signal-to-interference ratio (SIR) for a target at distance $d$ due to power transferred by an interferer at distance $r$ to a victim radar can be determined as follows. The useful signal power (the peak of the periodogram) and the interference power are 
\begin{align}
    S = |\gamma|^2 G^2_{\text{p}}, ~    I  \le  f |\gamma_{\rm{int}}|^2 G_{\text{p}}G_{\text{I}}~,
\end{align}
where $G_{\text{I}} \in [1,G_{\text{p}}]$ depending on the level of coherence of the interference\footnote{Level of coherence can be characterized through signal-to-interference mitigation gain, which is a function of FMCW waveform parameters of victim and interfering radars \cite{FMCW_int_SI_2014,p2p_int_fmcw_2018}.} (i.e., $G_{\text{I}}=1$ for incoherent interference, $G_{\text{I}}=G_{\text{p}}$ for coherent interference). 
Hence,
\begin{align}
\text{SIR}\ge \frac{|\gamma|^2 G^2_{\text{p}}}{f |\gamma_{\rm{int}}|^2 G_{\text{p}}G_{\text{I}}} = \frac{\sigma r^2}{(4\pi)d^{4}}
\frac{G_{\text{p}} B}{u \alpha \Bmax G_{\text{I}}}.
\end{align}
The first factor is out of the designer's control, while the second factor can be optimized  (via the duty cycle (small $u$), chirp slope (small $\Bmax/B$), radar FOV (thereby reducing $f$), effective processing gain (increase $G_{\text{p}}/G_{\text{I}}$)) to make sure that the SIR is much large than 1. Our results indicate that incoherent interference leads to a significant increase in the noise floor (tens of dBs), so it can reduce the ability to detect weak targets. Nevertheless, for nearby targets or targets with a high RCS, the SIR margin is sufficient to allow reliable detection. When interference is partially coherent this margin drops significantly. 

\section*{Interference Mitigation Strategies}
The impact of interference ranges from ghost targets and increases in noise floor. Both are detrimental to radar operations. Approaches to deal with interference can be grouped as either  \textit{reactive}, which aim to reduce the impact of interference after it has occurred, or  \textit{proactive}, which aims to avoid or reduce interference by design. We will describe various reactive strategies as well as three such proactive strategies: quasi-orthogonal FMCW waveforms; low-rate data communication between radar transmitters; an OFDM radar approach. 


\subsection*{Standard (Reactive) Approaches} 
Extensive studies were conducted in the context of the EU MOSARIM project, where a broad range of time-domain or frequency-domain signal processing techniques were proposed to mitigate FMCW and pulsed radar interference. These techniques are capable of deleting instantaneous interference that exists for a limited time or bandlimited interference that pollutes a specific portion of the whole radar band; while no solution is offered for the worst case recurring or wideband interference~\cite{MosarimFinalReport}.
The current attitude toward interference mitigation in the industry focuses on various techniques, including pulse to pulse processing and removing polluted pulses, sniffing and avoiding used frequencies, using frequency diversity, using narrow main beams or side-lobe null steering \cite{RadarCongestionStudy}.
These techniques are generally reactive strategies, which focus on getting rid of interference after it occurred, making it infeasible for highly-dynamic VANETs which require ultra-low latencies. Other reactive strategies exploit sparsity of useful signal and interference
components in different transform domains, namely, the DFT and short-time Fourier transform (STFT) domains, respectively, to extract the desired signal component \cite{sync_async_int_2019} or solve a sparse recovery problem to reconstruct the intervals in the range spectrum spoiled by interference \cite{Sparse_Samp_2017}. 
Interference avoidance techniques can also be more invasive, such as notifying the driver, disabling the sensor feature, shifting function to another sensor, reducing CFAR detection sensitivity \cite{RadarCongestionStudy}. However, these avoidance mechanisms either decrease the radar detection performance or disable the radar completely. 


\subsection*{Quasi-Orthogonal Waveforms} 
\subsubsection*{Concept} 

From the interference analysis, we established that the interference is proportional to $f=u \alpha \taumax/B$, where $u$ is the radar duty cycle, $\alpha$ is the chirp slope, $\taumax$ is the maximum target round-trip time, and $B$ is the radar bandwidth. Hence, by decreasing the chirp slope, or, equivalently, increasing the chirp duration, interference can be mitigated. A chirp $x(t)$ from \eqref{eq_base2} and a delayed chirp $x(t-\Delta\tau)$, repeated with period $T$, have power leakage/coupling 
\begin{align}
    C(\Delta\tau)& =\absbig{ {\frac{1}{{T}}\int_{0}^{T} e^{j \left[ \varphi(t) -  \varphi(t-\Delta\tau) \right]} \, \dd  t } }^2
	\le \frac{1}{{{\pi ^2}{B^2}{{ {{(\Delta\tau)}} }^2}}}	\label{eq:HH2}
\end{align}
for $\Delta\tau \neq 0$.  Since $B\Delta\tau/T$ is the instantaneous frequency of the chirp,  the coupling is the same as two sinusoids with frequency difference $B\Delta\tau/T$. Hence, we say that these waveforms are \emph{quasi-orthogonal}. As $\Delta\tau$ was arbitrary, this property is maintained under random starting and arrival times of FMCW waveforms. 
From \eqref{eq:HH2}, it is possible to derive the number of signals $N$ that cause acceptable interference, i.e., smaller than the power backscattered from a typical target. Suppose the chirp duration $T$ is divided into $N$ segments of duration $\Delta \tau$, so $\Delta \tau=T/N$, then, using \eqref{eq:interferencePower}--\eqref{eq:usefulPower}:
\begin{align}
   & P\frac{G\sigma\lambda^{2}}{(4\pi)^{3}d^{4}} \ge P C(\Delta \tau) \frac{G \lambda^2 }{(4\pi)^2 r^2} =P \frac{N^2}{(\pi B T)^2} \frac{G \lambda^2 }{(4\pi)^2 r^2}\nonumber
     \Rightarrow N \le TB \frac{\sqrt{\sigma}}{r} 
\end{align}
if $r$ and $d$ take on similar values. For $r=500 ~\text{m}$, $B=1~\text{GHz}$, $\sigma  = 10~\text{m}^2$, using one long chirp of duration $T=10~\text{ms}$ leads to over 60,000 quasi-orthogonal waveforms. 
 The challenges of retrieving velocity and range data as well as physically realizing such a radar is now briefly described.

\subsubsection*{Signal Processing} 
\begin{figure} 
	\centering
	\includegraphics[width=0.5\linewidth]{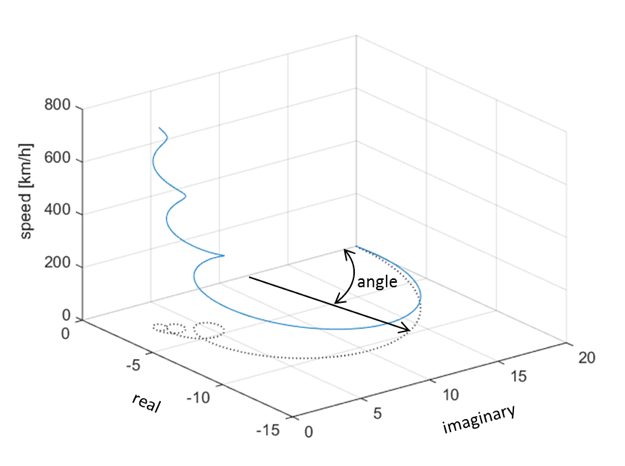}
	\vspace{-0.5cm}
	\caption{Retrieval of velocity from integral in \eqref{eq:HH7}. The integral as a complex valued function of speed (in km/h) is the blue line trace. A radius vector for the complex integral is drawn selecting the origin so to make this vector meet the trace in an as perpendicular fashion as possible. The angle of this vector thereby sensitively varies with integral value. Tabulating the angles for varying velocity, the latter are readily found from integral angle values by an inverse look up procedure within the significant speed interval in which the angle-speed relation remains 1-1. \label{fig:HH1}}
	\vspace{-0.5cm}
\end{figure}

While in \eqref{eq_ykn} speed and range appear in an ambiguous combination in a single chirp, this expression is only an approximate representation of Doppler shift, as we neglected several constants and small terms, which are negligible for small $T$. 
For long chirps, these neglected values should be considered, so that 
the range speed ambiguity can be resolved  within a single chirp. 
Formally, 
\begin{align} \label{eq_rt_fmcwHH}
y(t) = r(t) \xconjt 
= \tildegamma \, e^{j 2\pi f_c \nu t} e^{-j 2\pi \alpha \tau t}\rect{T}{{t - \tau}}
e^{j\varphi_0} e^{j2 \pi t^2 \nu \alpha}   + w(t) \xconjt 
\end{align}
where $r(t) = r_0(t)$ with $r_k(t)$ being defined in \eqref{eq_rt_fmcw2}, $\tildegamma$ is a real quantity denoting the target reflectivity, and $\varphi_0  = \pi \alpha \tau( {\tau - {{{2f_c}}}/{\alpha}} )$ is an absolute phase. 
The Fourier transform $Y(f)$ of $y(t)$ will have a maximum at $f^*=f_c \nu -\alpha \tau $, with 
\begin{align}
    Y(f^*)=\tildegamma e ^{j\varphi_0 }\int_{0}^{T}e^{j 2 \pi t^2 \nu \alpha}\, \dd t~.
    \label{eq:HH7}
\end{align}
Hence, finding the maximum of $|Y(f)|^2$ yields an estimate of $f_c \nu -\alpha \tau$, which allows us to write 
\begin{align}
	\varphi_0  = \pi \frac{f_c \nu -f^*}{\alpha}( {{f_c (\nu-2) -f^*}} ) ~.
	\label{eq:HH8}
\end{align}
We can then invert the expression \eqref{eq:HH7} to solve for $\nu=2 v / c$. 
The solution has to be found numerically, making use of the fact that the target reflectivity $\tildegamma$ is real and positive and that the complex angle of the integral increases monotonically with velocity within a significant velocity span. The task of retrieving velocities under these circumstances is illustrated in Fig.~\ref{fig:HH1}.

 \begin{figure} 
 	\centering
 	\includegraphics[width=0.5\linewidth]{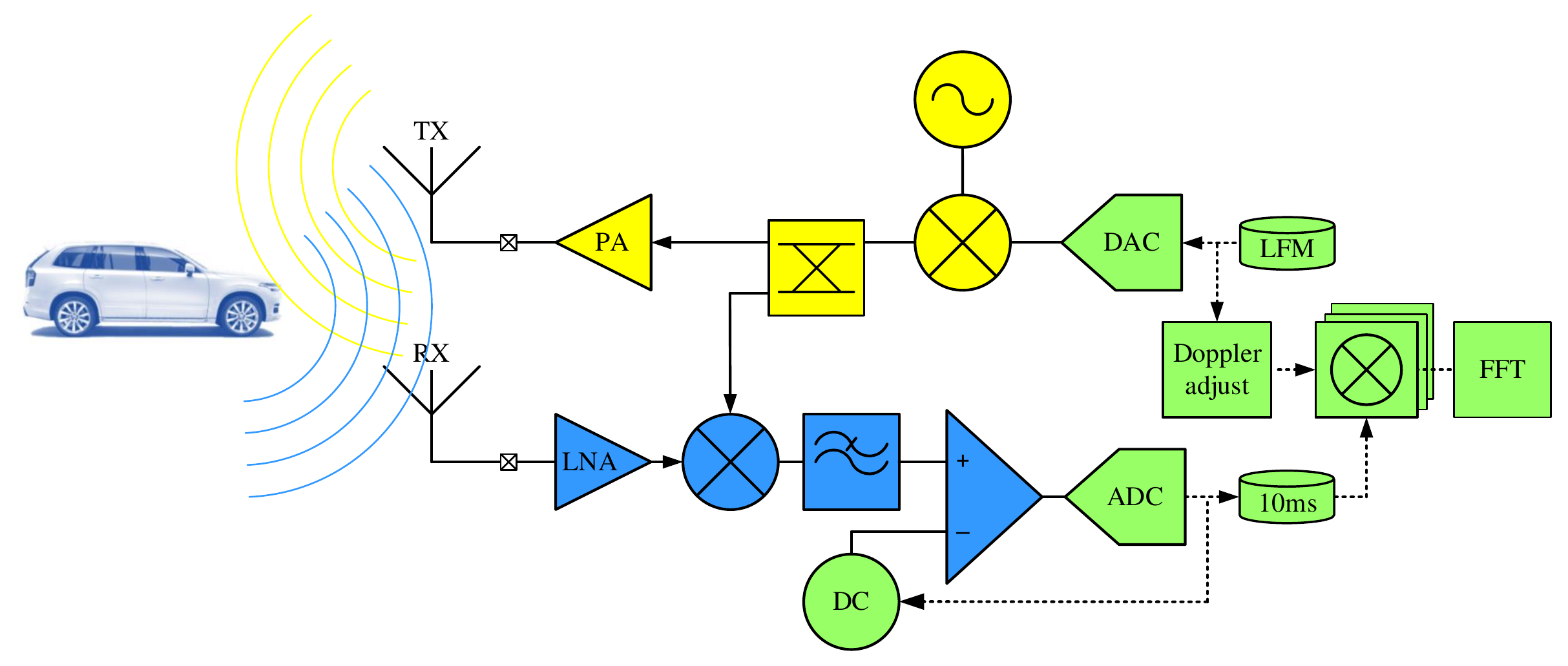}
 	\caption{Simplified scheme of a single slow sweep FMCW radar allowing large numbers of orthogonal signals. The LFM waveform is obtained digitally at baseband and is upconverted in the analog domain. This upconverted signal is mixed with the receive signal yielding downconversion to a signal with bandwidth essentially set by 50 MHz. The dominant power in the downconverted signal will come from transmit signal which however appears as a DC component. This is removed by a DC canceller stage after which the signal is digitized. The canceller operates on the principle of minimizing at ADC output at some rate much slower than the sweep time and has the vital function of avoiding ADC saturation. In the digital domain further mixing with Doppler offsets may be required to cover a very wide range of target speeds. \label{fig:HH2}}
 \end{figure}


Comparing the outlined approach to that of the slow/fast time
FMCW radar case above, note the principal difference that presently velocity
is determined by measuring complex amplitude. For pulse-Doppler radar, velocity is found by locating the target peak in the Doppler spectrum. The present approach assumes that just one target is present for any beat frequency whereas pulse-Doppler radar may handle several targets within the same range/Doppler resolution cell. On the other hand the attained beat frequency resolution is refined in proportion to the prolonged sweep. In effect both methods
exhibit the same low probability that two separate targets should be superimposed
the same resolution bin.
The single slow sweep method for obtaining nearly orthogonal waveforms contains a further challenge compared to pulse-Doppler  radar, apart from the novel signal processing required. Indeed the steep ramps of pulse-Doppler FMCW frequency offsets the target response as compared to the transmit signal. Presently, the radar is based on subtraction leakage appearing as DC component rather than filtering with respect to range and Doppler frequency offsets. The conceived radar scheme in Fig.~\ref{fig:HH2} indicates that \eqref{eq:HH7} can be slightly modified by digitally modifying the signal phase by some offset,
 thereby changing the unambiguous speed range for \eqref{eq:HH7}. To effectively cover a sufficiently large velocity span, two or three such velocity channels should be processed in parallel. The overall processing burden still remains fully reasonable.

\subsubsection*{Implications}
The freedom of multiple orthogonal radar channels can be brought into practice in different ways. The considered case with a very large number of channels allows for the convenient method of simply not requiring any common scheduling of the channels adopted (apart from the several radars which may be located in the same vehicle in which case channel coordination is simple to achieve). The number of vehicles in such proximity to each other that an interference conflict is imminent will be much smaller than the number of available channels. Just selecting the radar channels randomly, chances are good that there will be no interference. In the case of interference, the individual vehicle will then randomly pick another channel and with high probability the conflict thereby is resolved. 

\subsection*{Coordinated Transmission via Wireless Communication}
\begin{figure}[ht!]
	\centering
	\vspace{-0.1cm}
	\includegraphics[width=0.6\linewidth]{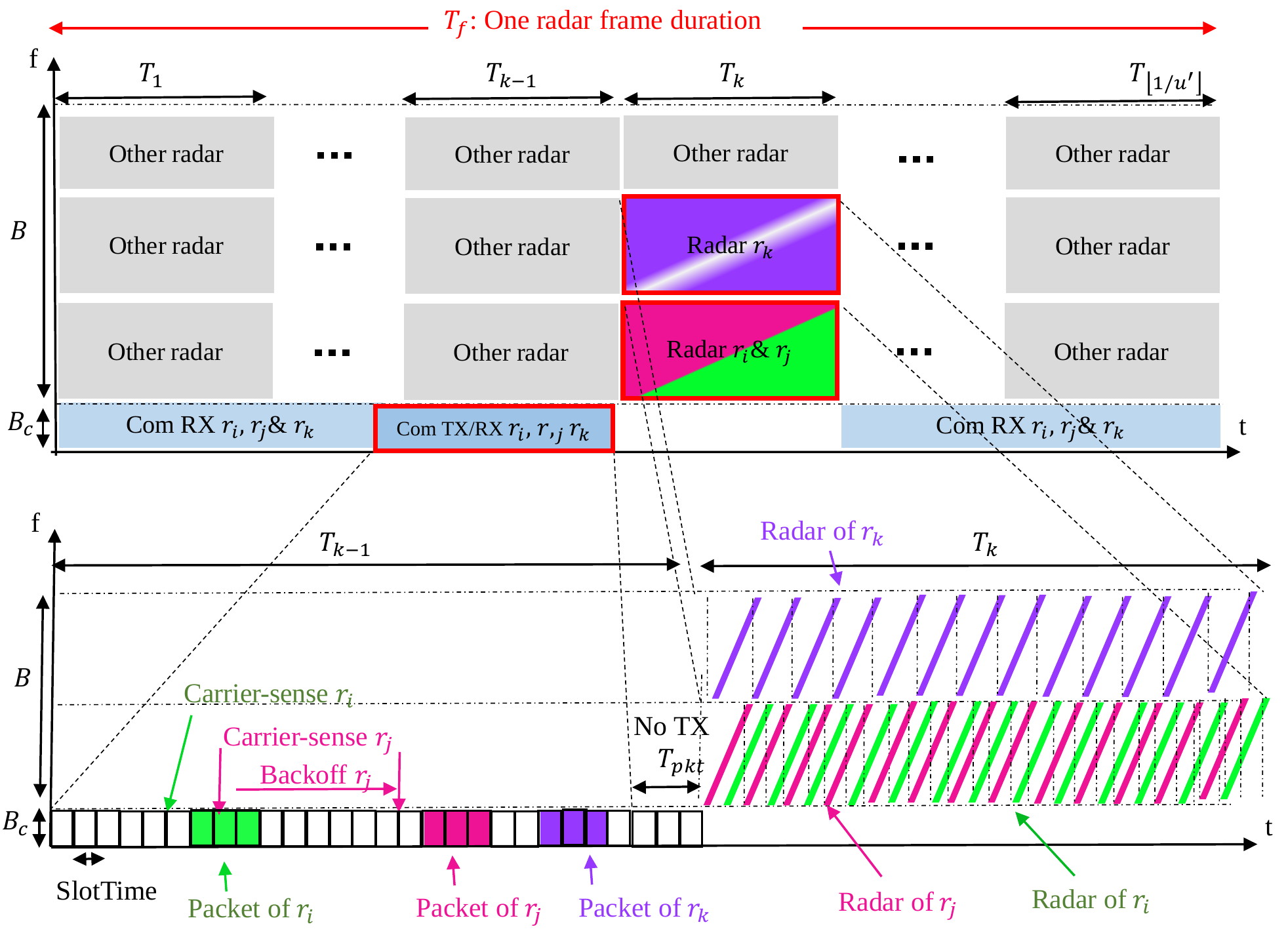}
	\vspace{-0.3cm}
	\caption{An example showing the sharing of frequency-time resources by three radar communication units $r_i$, $r_j$ and $r_k$. The whole bandwidth is divided into a communication channel of bandwidth $B_c$ and sweep bandwidth $B$, where various types of automotive radars might use portions of $B$. One radar frame $T_f$ is divided into time slots $T_k, 1 \le k\le \lfloor 1/u'  \rfloor$, where $u'=(N+1)T/T_f$ is the modified radar duty cycle. The radar frame is divided into time slots, where radar transmission/reception takes place in $T_k$, communication transmission is done prior to radar at $T_{k-1}$ and reception at anytime except $T_k$. Each RCU broadcasts it packet over the communication channel through CSMA, which includes the starting time and frequency band information of its radar transmission. In this example interference is resolved because $r_i$ and $r_j$ are allocated to quasi-orthogonal time slots, whereas $r_k$ is allocated to a different frequency band. Any conflicts in frequency-time resource sharing that occur in one frame $T_f$ are resolved in the following frames.}
	\label{fig_RadChat}
\end{figure}

Another way to mitigate interference is coordination of automotive radars through communications so that radars are assigned disjoint frequency-time-space resources, making use to the fact that (i) radars use only a small fraction $\Bmax/B$ of the available bandwidth; (ii) radars are only active a fraction $u$ of the time; (iii) radar signals are blocked  --mostly by other vehicles-- and limited by the field of view. 
The assignment requires coordination for allocation of frequency-time resources through distributed network communication. Such communication can be achieved either via a dedicated technology, such as 802.11p or cellular V2X (C-V2X) communication. Alternatively, one can exploit the similarity of the radar circuitry to standard communication hardware, and upgrade automotive radars  to joint radar communication units (RCU), which use the same hardware for both radar and communications and are composed of data link and physical layers. The radar and communication functionality are time-multiplexed, where communication occurs over a fixed and dedicated communication bandwidth (with bandwidth limited by the ADC), which is free of radar transmissions. The time-multiplexing is possible because of the idle period. Hence, when $u\approx 1$, RCUs cannot be used and 802.11p or C-V2X are more appropriate. Nevertheless, we will consider here the RCU approach, as it is readily modified when using another dedicated communication technology. As is typical in VANETs, communication is unacknowledged with a distributed MAC based carrier sense multiple access (CSMA). The goal of the communication is to assign radars to time slots, so that different radars remain quasi-orthogonal. In contrast to the long chirps described previously, we consider standard short chirps, thus limiting the number of transmissions per chirp period. 
The frequency-time resources are shared as illustrated in Fig.~\ref{fig_RadChat} for three RCUs $r_i$, $r_j$ and $r_k$.  The basic principle is as follows. Each vehicle initially assigns starting times to the automotive RCUs mounted on this vehicle through a central processor. These are broadcast to neighbouring vehicles during a communication slot. All RCUs on a vehicle broadcast short control communication packets at the same time over the communication band. 
The broadcast communication packet includes information about the starting times used by all RCUs on that vehicle. Other RCUs or vehicles, which receive this information, store this information in a database and allocate themselves non-overlapping starting times based on the stored information. A priority index is used to prioritize the dissemination of the resource allocation of a larger group of vehicles in order to avoid fluctuations in the distributed VANET. 
\begin{figure}[ht!]
	\centering
	\includegraphics[width=0.5\linewidth]{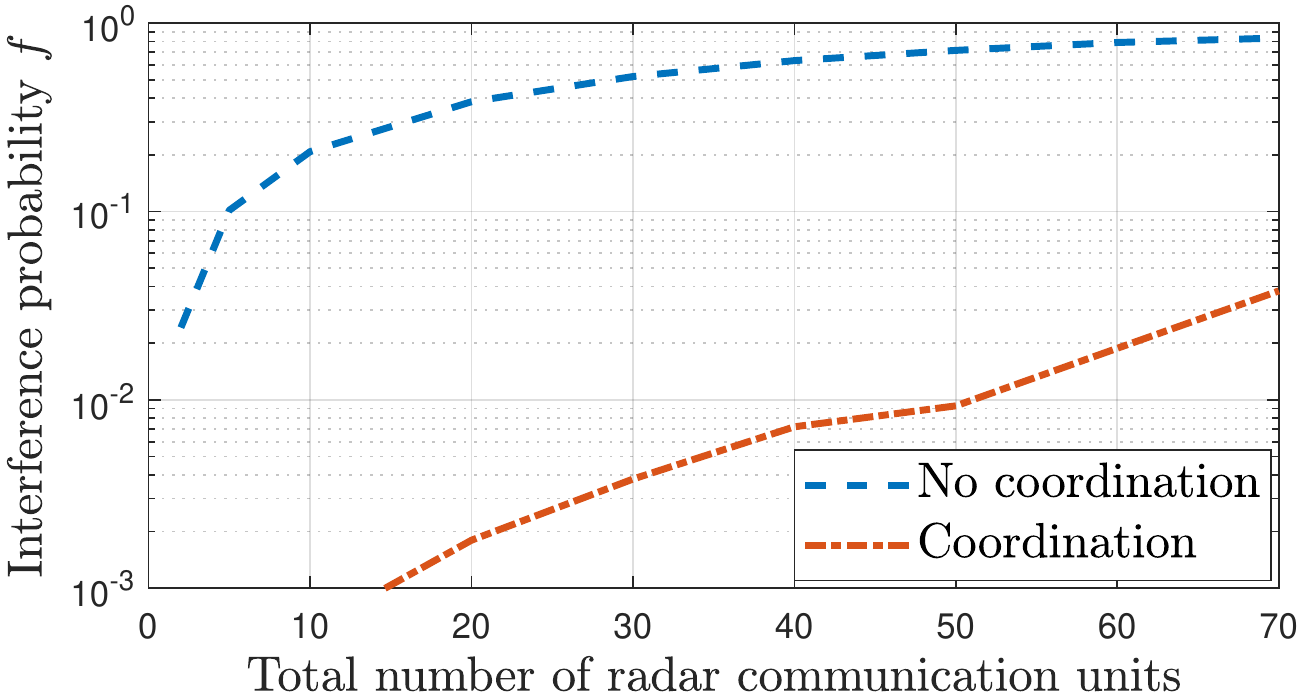}
	\vspace{-0.3cm}
	\caption{Radar interference probability $f$ after one frame time (2ms) with coordinated transmission for varying number of radars for 10,000 Monte-Carlo simulations, $B=\SI{1}{\giga\hertz}$, $B_\text{max}=\SI{50}{\mega\hertz}$, $T=20\mu s$, $f_c=\SI{79}{\giga\hertz}$, $K=99$.}
	\label{fig_RadChat_result}
	\vspace{-0.5cm}
\end{figure}

A practical implementation requires synchronization among radars, which can be achieved through GPS or via a dedicated synchronization protocol. With reasonable synchronization requirements (around 1--2 $\mu$s error), \cite{CananITS} demonstrated significant reductions in radar interference within a few tens of ms. Fig.~\ref{fig_RadChat_result}  shows the interference probability (i.e., the expected value of $f$ from \eqref{eq:interferencePower}) as a function of the number of interfering radars, with and without coordination. 
The potential of such protocols to adapt radar signals according to changing traffic conditions via communications offers intelligent radar sensing strategies, such as cooperative localization, disabling unnecessary sensing, etc. 


\subsection*{Joint Radar and Communication}

\begin{figure*}
	\vspace{-0.5cm}
	\begin{center}		
		\subfigure[]{
			\label{fig_sf_ofdm}
			\includegraphics[scale=0.15]{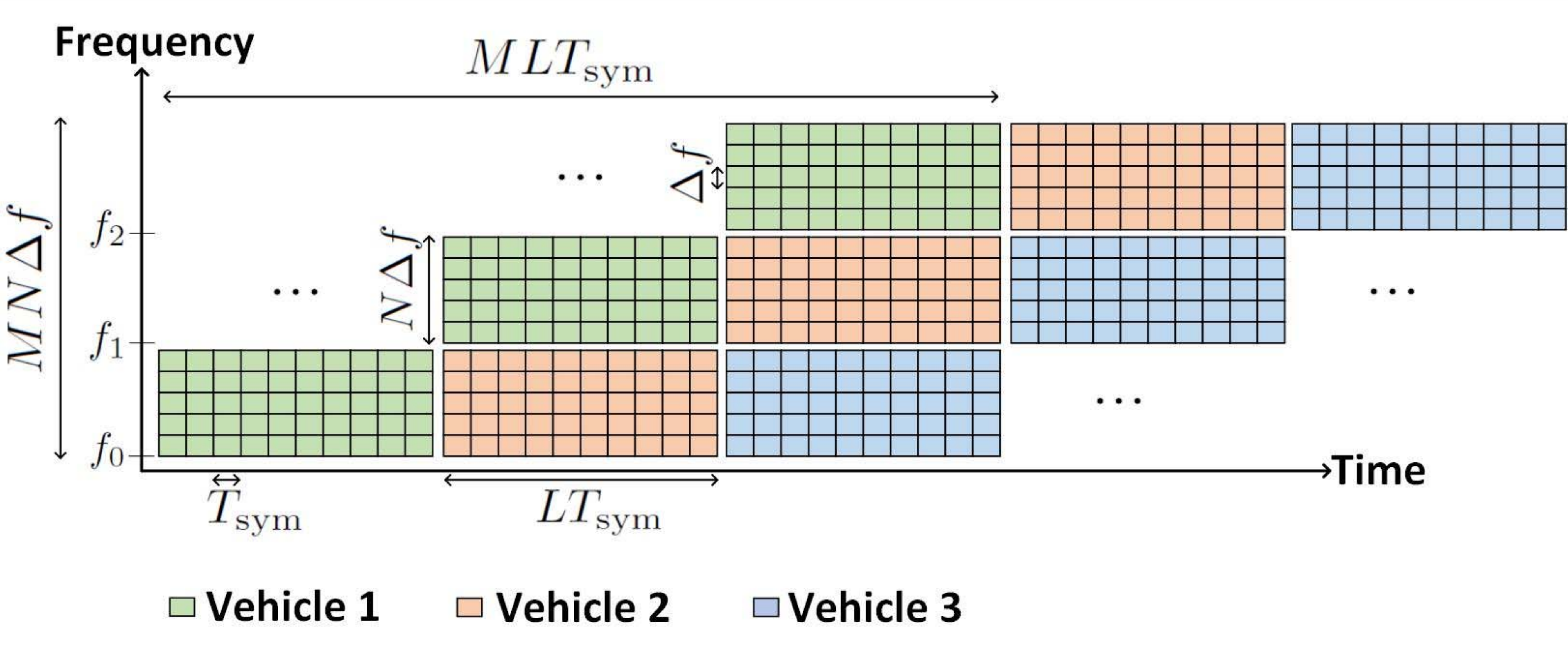}
		}
		\hspace{-0.2cm}
		\subfigure[]{
			\label{fig_n_ofdm}
			\includegraphics[scale=0.15]{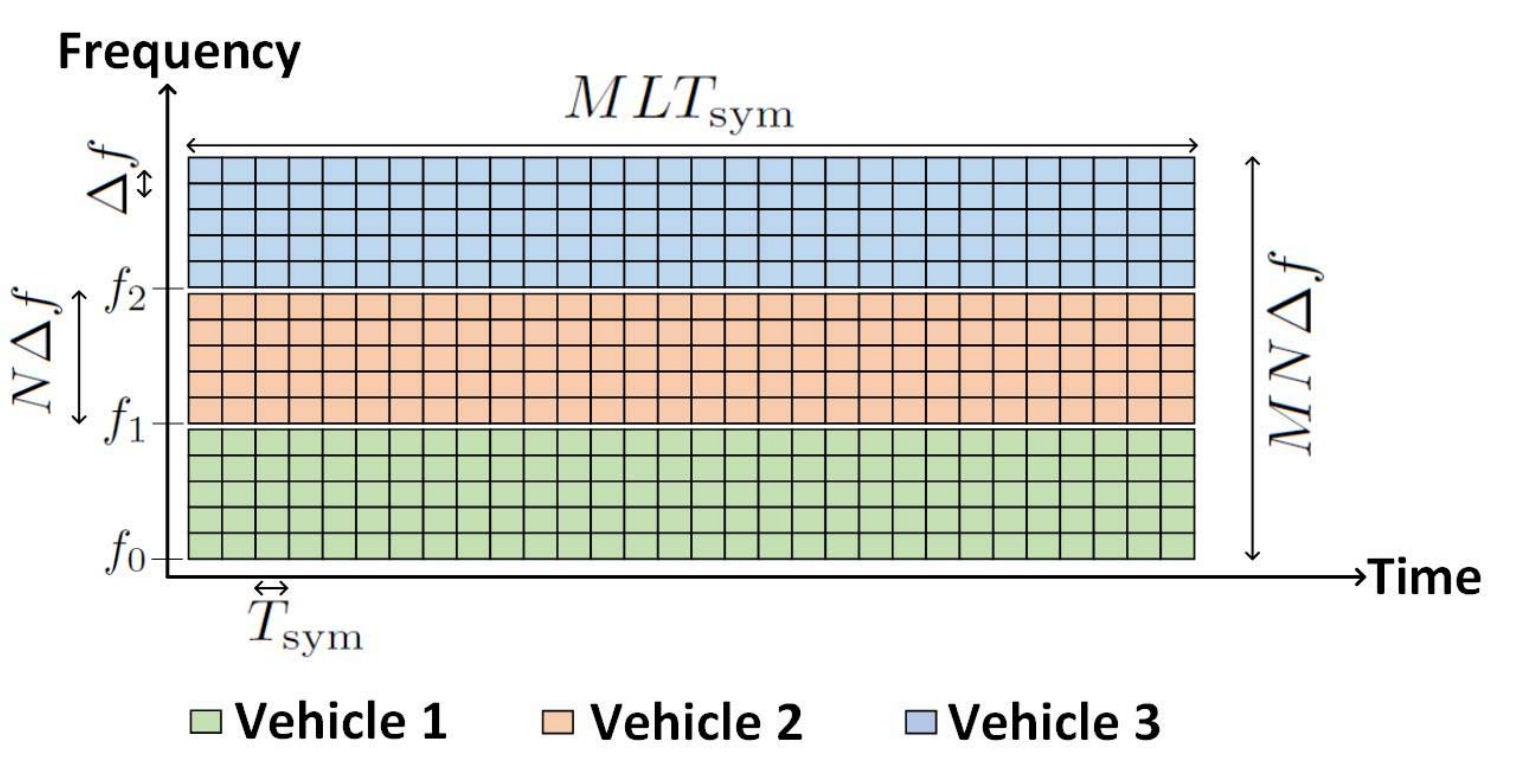}
		}
	\end{center}
	\vspace{-0.5cm}
	\caption{Time-frequency resource coordination in a joint radar communications vehicular network for \subref{fig_sf_ofdm} stepped-frequency OFDM 
	and \subref{fig_n_ofdm} narrowband OFDM.}
	\label{fig_ofdm_resource}
	\vspace{-0.5cm}
\end{figure*}

A third, more forward looking alternative is to exploit the fact that radar and communication  systems operate in similar frequencies, and develop a system that can perform the dual role of radar and communication  \cite{Bs_radar_2019}, coined RadCom. Both pilot and data from the transmitted signal can be exploited for radar functions when processing the backscattered communication signal.  
A prominent candidate for this is OFDM, which is the de facto  waveform for all cellular and wifi-based standards, due to its flexibility and robustness to wireless propagation effects. OFDM has also been studied extensively as a radar waveform \cite{RadCom_Proc_IEEE_2011,Firat_OFDM_2012,OFDM_Radar_Phd_2014,SPM_JRC_2019,Passive_OFDM_2010}, but is limited by the ADC bandwidth, which is generally orders of magnitude smaller than the radar bandwidth, which in turn limits radar resolution. 
A way around this problem is the use of \emph{Stepped-Frequency OFDM}, which involves consecutive OFDM frames, each transmitted with a different carrier frequency \cite{Stepped_OFDM_TAES_2015,Stepped_OFDM_CS_2018,Stepped_Carrier_OFDM_2018}. The main rationale behind the use of stepped-frequency OFDM as a RadCom waveform is to surpass the range resolution limitation of conventional OFDM radar (which is imposed by ADC bandwidth) via frequency hopping across individual OFDM frames with low baseband bandwidth \cite{Stepped_Carrier_OFDM_2018,Stepped_OFDM_CS_2018}, while maintaining standard wideband OFDM as a special case. 
Fig.~\ref{fig_sf_ofdm} illustrates an exemplary time-frequency plot of a stepped-frequency OFDM waveform. To avoid interference, different vehicles are assigned orthogonal resources, as shown for 3 vehicles. 
Hence, the stepped-frequency OFDM can exploit high resolution offered by the \textit{large total bandwidth} $M N \deltaf$ by joint processing of $M$ individual OFDM frames on different carriers while simultaneously requiring a low-rate ADC to sample \textit{small baseband bandwidth} ($N \deltaf$) OFDM blocks. For each carrier $L$ OFDM symbols of duration $\Tsym$ are sent, constituting a frame. The choice of $N$, $M$, $L$, $\Delta f$, and the hopping pattern provides flexibility in the RadCom waveform, and enables us to provide radar performance similar to an equivalent wideband OFDM radar, but with low-rate, {low-cost ADCs} \cite{Stepped_OFDM_CS_2018}.  

\begin{figure*}
	\vspace{-0.2cm}
	\begin{center}		
		\subfigure[]{
			\label{fig_nveh_snr}
			\includegraphics[scale=0.3]{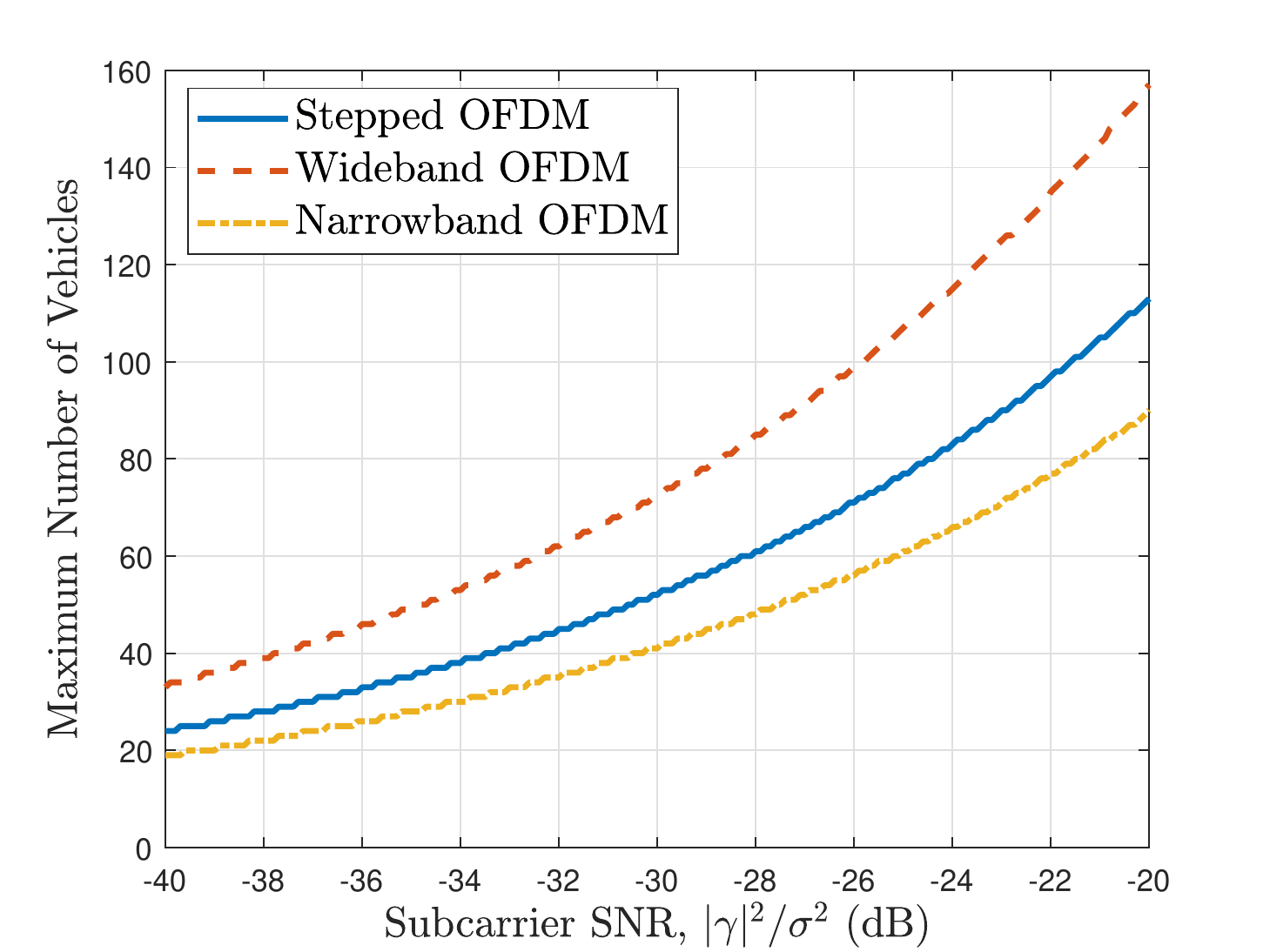}
		}
		\hspace{-0.2cm}
		\subfigure[]{
			\label{fig_nveh_range}
			\includegraphics[scale=0.3]{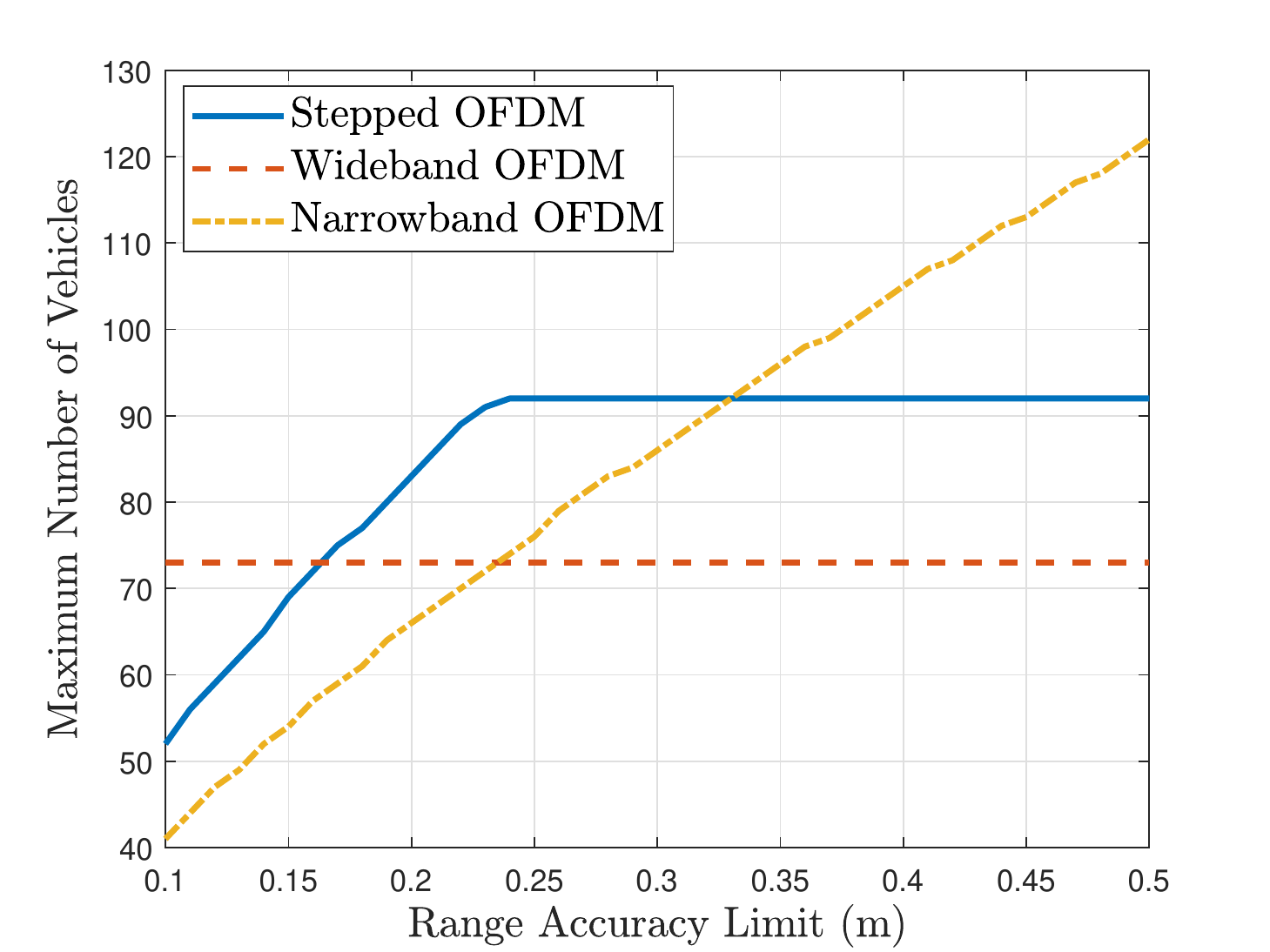}
		}
		\hspace{-0.2cm}
		\subfigure[]{
			\label{fig_nveh_vel}
			\includegraphics[scale=0.3]{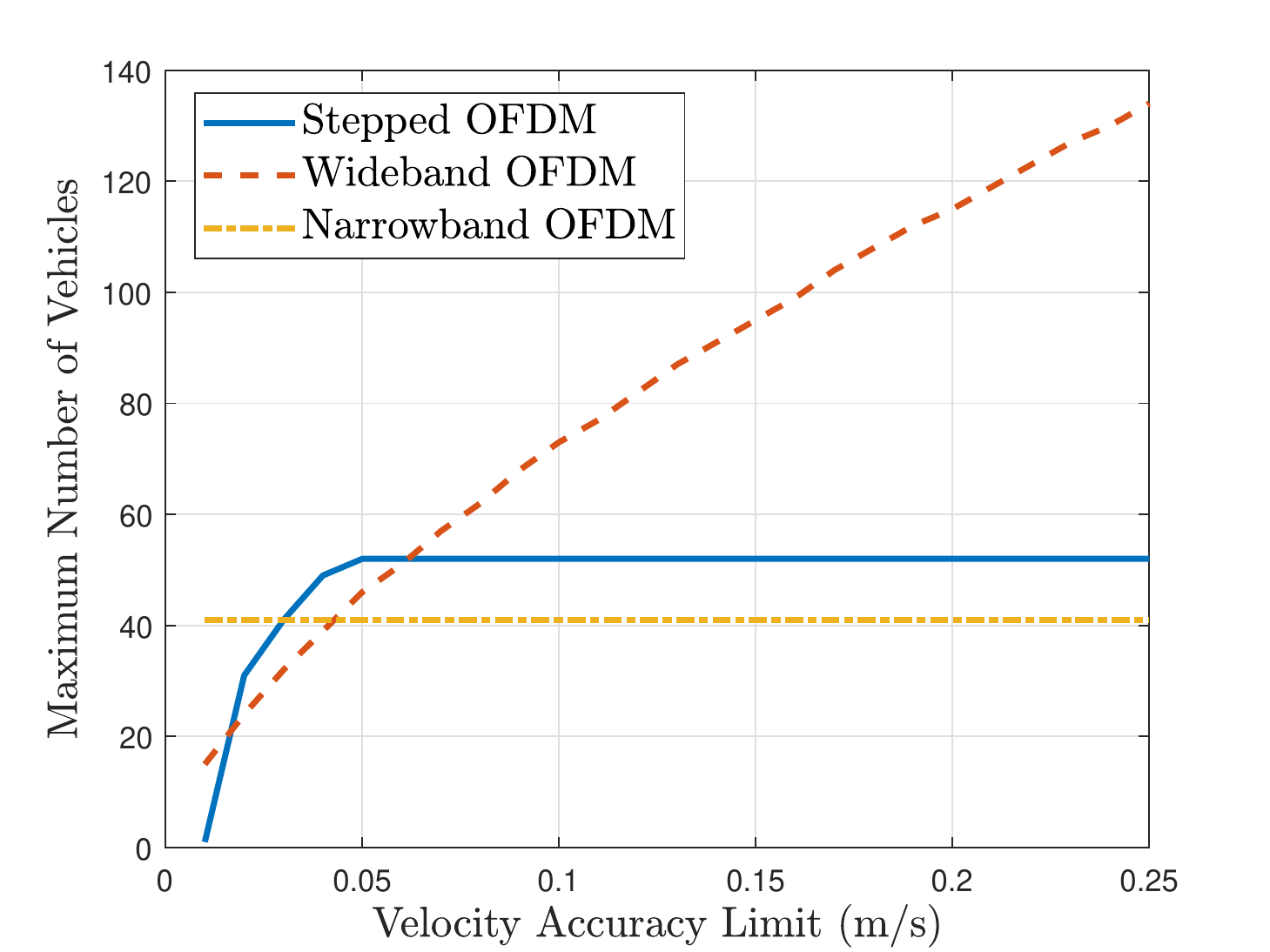}
		}
	\end{center}
	\vspace{-0.5cm}
	\caption{Maximum number of vehicles that can be accommodated within a time-frequency resource defined by $\Btot = 1 \, \rm{GHz}$ and $\Ttot = 30 \, \rm{ms}$ for the three different OFDM waveforms, where $f_0 = 77 \, \rm{GHz}$, $\deltaf = 500 \, \rm{KHz}$, $\Tcp = 400 \, \rm{ns}$. Stepped OFDM and narrowband OFDM schemes require an ADC with $50 \, \rm{MHz}$ rate, whereas wideband OFDM requires $1 \, \rm{GHz}$-rate ADC. \subref{fig_nveh_snr} With respect to subcarrier SNR, where range and velocity accuracy limits are $0.1 \, \rm{m}$ and $0.1 \, \rm{m/s}$, respectively. \subref{fig_nveh_range} With respect to range accuracy limit, where subcarrier SNR is $-30 \, \rm{dB}$ and velocity accuracy limit is  $0.1 \, \rm{m/s}$. \subref{fig_nveh_vel} With respect to velocity accuracy limit, where subcarrier SNR is $-30 \, \rm{dB}$ and range accuracy limit is  $0.1 \, \rm{m}$.}
	\label{fig_ofdm_results}
	\vspace{-0.5cm}
\end{figure*}

\subsubsection*{Signal Processing and Resource Allocation}
Under standard assumptions (cyclic prefix longer than $\tau$ \cite{Firat_OFDM_2012,OFDM_Radar_Phd_2014,SPM_JRC_2019} and small Doppler approximation \cite{Passive_OFDM_2010}), the received symbol on the $n$th subcarrier for the $\ell$th symbol of the $m$th frame can be written as (considering the same radar environment with a single target as specified in \eqref{eq_rt_fmcw2}) \cite{Stepped_Carrier_OFDM_2018}
\begin{align} \label{eq_rec_final}
\ymln = & \gamma \, \xmln \,  e^{- j 2 \pi (f_m + n \deltaf) \tau }e^{j 2 \pi f_0  (mL + \ell+1) \Tsym  \nu  } + \wmln
\end{align}
where $\xmln$ denotes the complex data or pilot symbol, $\gamma$ is the complex channel gain and $\wmln$ is the additive noise term with variance $\sigma^2$. Delay estimation can be performed by matched filtering the data cube $\ymln$ across frame-frequency dimensions ($m$ and $n$ directions), while processing along frame-time dimensions ($m$ and $\ell$ directions) can provide an estimate of Doppler. Frequency hopping across consecutive OFDM frames introduces \textit{delay-Doppler coupling}, which can be overcome by incorporating phase correction terms in the DFT implementation of matched filtering \cite{Stepped_Carrier_OFDM_2018}.

Time-frequency resource allocation scheme coordinated by a central unit (e.g., a 5G base station) helps to alleviate mutual interference among radars on different vehicles, similar to conventional OFDM radar networks \cite[Ch.~4]{OFDM_Radar_Phd_2014}. 
Resources can be assigned to maximize the number of vehicles that can be fit into a given time-frequency block such that each vehicle meets preset radar accuracy requirements\footnote{To characterize radar accuracy, we employ the Cram\'{e}r-Rao bound (CRB) \cite{OFDM_Radar_Phd_2014} on variances of unbiased estimates of delay and Doppler parameters using the signal model in \eqref{eq_rec_final}.}. Fig.~\ref{fig_ofdm_results} shows exemplary results for the three different OFDM schemes (stepped-frequency, narrowband, and wideband, where in the latter case each vehicle uses the total bandwidth (and thus requires an ADC with $1 \, \rm{GHz}$ sampling rate) for a duration $L\Tsym$ and then remains silent for a duration $(M-1)L\Tsym$). 
As seen from the figures, the stepped-frequency OFDM radar can support more vehicles in a given spectral resource than the conventional narrowband OFDM radar with the same hardware requirements since the former offers the flexibility to trade off a decrease in Doppler accuracy for an improvement in ranging accuracy (frequency hopping increases ranging accuracy and reduces Doppler accuracy). In Fig.~\ref{fig_nveh_range}, as wideband OFDM is essentially limited by the velocity accuracy constraint, relaxing the range accuracy constraint does not further improve its performance (similarly, for narrowband OFDM in Fig.~\ref{fig_nveh_vel}).

As a final remark, we note that the stepped OFDM provides a design trade-off between the narrowband and wideband OFDM schemes, retaining the improved resolution and accuracy properties of the wideband OFDM with much reduced hardware requirements as in the case of narrowband OFDM.


\section*{Outlook and Challenges}
In this section, we consider the main research and development challenges for the coming years. For communication-based interference mitigation strategies, the coexistence between radar and communication signals is an important challenge. For joint radar and communication signals, there is a potential of a revolution of cellular-type signals (e.g., 5G NR) to be reused for radar purposes \cite{Bs_radar_2019}, opening exciting synergies and avoiding the need for dedicated RF hardware all-together. The extended frequency bands made available for 5G are interesting by themselves due to the possible improvement in radar resolution, which, together with the already standardized orthogonal signaling, establishes an exciting area for automotive sensing. 
The main challenge is to find solutions that will enable radar and communication functionalities with such a low information latency that vehicle safety is not compromised in any traffic scenario. These solutions should include techniques for a fair distribution of the available time and frequency space for all users. It must also secure a low data loss for both radar and communication which, of course, is the aim of minimizing the possibility of interference.

For the generation and detection of slow chirps, new hardware architectures will be needed. It will push a migration from analog towards digital electronics and signal processing, which will pave the way for technologies such as imaging radar. The other modulation waveforms proposed in this paper will also require hardware that differs from the current radar designs. The analog to digital, and vice versa, conversion will be close to the RF front end making more complex digitally generated and filtered waveforms possible.
To push this development forward, we will implement a demonstrator platform, complete with millimeter wave front ends, high speed digital signal generation and signal acquisition, and independent generation of arbitrary interference. The outlined methods in this paper will be tested and evaluated on the demonstrator platform in a realistic environment. The intention is to use this demonstrator to verify the theoretical analysis regarding interference probability and SINR for different types of modulation and verify the speed measurement method for the slow ramp modulation.

Further development would include the integration of critical electronic components in CMOS technology, to ensure that a complete solution is feasible to implement in commercial scale. Advanced CMOS technologies can also facilitate the implementation of alternative waveforms on automotive radars, such as phase modulated continuous wave (PMCW). 
Compared to the widely used FMCW radar, the PMCW waveform has the major disadvantage of requiring very high-rate ADCs to sample wideband code sequences. 
On the other hand, it possesses several advantages making it attractive for future deployments, including improved robustness to interference via proper code design,  
not requiring a highly linear frequency ramp synthesizer, and inherent applicability to MIMO radar configurations through code orthogonality across multiple antennas. 
From the perspective of radar interference mitigation and radar communications convergence, we expect that the main focus of the automotive industry in the coming years will be on the cost and integration of both analog and digital functions on the same silicon chip to reduce the likelihood of hardware failure.

Vehicle radars can also be expected to operate in higher frequency bands, 100--300 GHz, to enable more bandwidth, cost reduction and miniaturization of the hardware. It is possible to influence the regulators to include some level of standardization in automotive radars, which is needed for mitigation of interference among different automobile brands~\cite{RadarCongestionStudy}, before new frequency spectrum is made available in the higher RF bands.
Hence, it is timely to conduct research and discuss the development challenges about automotive radar interference before it becomes a problem. 
\vspace{-0.3cm}


\bibliographystyle{IEEEtran}
\bibliography{references_spm.bib}

\end{document}